\documentclass[a4paper,12pt]{article}
\usepackage{a4wide}
\usepackage{graphicx}
\usepackage{amsmath}
\usepackage{cite}
\usepackage{subfigure}

\newcommand{\ba}{\begin{eqnarray}}
\newcommand{\ea}{\end{eqnarray}}
\newcommand{\be}{\begin{equation}}
\newcommand{\ee}{\end{equation}}

\newcommand{\DS}[1]{/\!\!\!#1}
\begin{document}
\begin{titlepage}
\begin{flushright}
SI-HEP-2005-08\\
IRB-TH-21/05
\end{flushright}
\vfill
\begin{center}
{\Large\bf Annihilation effects 
in \(B\to\pi\pi\)\\[0.3cm] from QCD Light-Cone Sum Rules}\\[2cm]
{\large\bf A.~Khodjamirian\,\(^1\), Th.~Mannel\,\(^1\), M.~Melcher\,\(^1\) and
B.~Meli\'c\,\(^2\)}\\[0.5cm]
{\it\(^1\) Theoretische Physik 1, Fachbereich Physik,
Universit\"at Siegen,\\ D-57068 Siegen, Germany \\
\(^2\) Rudjer Bo\v skovi\'c Institute, Theoretical Physics Division,\\
HR-10002 Zagreb, Croatia}
\end{center}
\vfill

\begin{abstract}
Using the method of QCD light-cone sum rules,
we calculate the $B \to \pi\pi$ hadronic matrix elements 
with annihilation topology. We obtain a finite result, 
including the related strong phase. Numerically, 
the annihilation effects  in $B\to \pi\pi$ turn out to 
be small with respect to the factorizable emission mechanism.
Our predictions, together with the earlier sum rule estimates of  
emission and penguin contributions, are used for the 
phenomenological analysis of $B\to \pi\pi$ channels.
We predict a $\Delta I=1/2$ transition amplitude which
significantly differs from this amplitude extracted 
from the current data.  
\end{abstract}
\vfill

\end{titlepage}

\section{Introduction}
Charmless $B$ decays play an increasing role as a possible window to physics
beyond Standard Model (for a review see. e.g., \cite{reviews}).  
In particular, the decays of the type $B \to \pi \pi$
and $B \to K \pi, K \bar{K}$ are actively investigated using various 
approaches \cite{BBNS1,BBNS2,PQCD,charmpeng,BuchallaSafir,Buras_etal,
SCET,FH,BBNSfits,Rosneretal,Grossman_etal}. 

One promising method to evaluate charmless nonleptionic $B$ decays is 
QCD factorization (QCDF) \cite{BBNS1,BBNS2}, which to leading order in $1/m_b$
and $\alpha_s (m_b)$ 
yields the well known formulae of naive factorization. However, even when 
QCD corrections and estimates for the penguin contributions are included,  
QCDF is only marginally consistent with the data for $B \to \pi \pi$. 
As a more general analysis of data 
shows,  one indeed needs 
additional contributions to the
decay amplitudes which correspond to the $\Delta I = 1/2$ piece of the
Hamiltonian. Comparing with
different topologies of 
the decay diagrams, 
this means an enhancement of either the penguin or annihilation 
contributions. 
Especially interesting is the  weak annihilation 
(or weak exchange) 
of the $b$ quark and the light antiquark in the $B$ meson. 
In QCDF, formally suppressed 
as $1/m_b$, the annihilation contribution is divergent.
Fits of QCDF to data \cite{BBNSfits}, with the annihilation part
replaced by finite parameters, yield large effects. Importantly,
the fits themselves cannot clearly distinguish between 
the annihilation and penguin mechanisms. 
In order to assess the relative 
importance of  the 
annihilation contribution, one needs  a different approach in QCD
which allows to calculate $B\to \pi\pi$ hadronic matrix elements 
with various topologies.

In the present paper we will  employ the technique 
suggested in \cite{AKBpipi}. It is  based on light-cone sum rules 
(LCSR) \cite{lcsr}, the method which adapts 
the general idea of QCD sum rules \cite{SVZ}
for  the amplitudes of exclusive hadronic processes. 
One well known application of LCSR   
is the calculation of the $B\to \pi$ form factor \cite{Bpi}.
In the sum rule approach, as opposed to QCDF
or to the perturbative hard-scattering QCD approach (PQCD)
\cite{PQCD}, one does not directly  represent the hadronic  
matrix element in terms of quark-gluon diagrams.
The QCD calculation takes place for 
the correlation function, a more general object,
where the separation of short and long-distance parts
into hard-scattering amplitudes and pion distribution 
amplitudes (DA's) respectively, is possible ``by construction'', that is, due
to an appropriate choice of 
kinematical variables. 
Both hard and soft gluon effects are systematically 
included in this calculation, contributing to different
terms of the light-cone operator-product expansion (OPE).
The hard gluon exchanges enter the hard-scattering amplitudes, 
whereas the soft gluon effects are  
represented by quark-antiquark-gluon DA's of the pion.
The hadronic matrix element appears  as a part of the hadronic 
dispersion relation for the correlation function.
Matching this relation to the result of the QCD calculation,
one employs quark-hadron duality
to separate the ground-state $B\to \pi\pi$ amplitude   
from the background of excited hadrons. 
The advantage of this technique, 
as will be explained in more details further,
is a possibility to associate different decay topologies 
in the hadronic matrix element
with the corresponding diagrams in
the OPE, hence the relative 
contributions of various operators and topologies 
to the decay amplitude  can be 
estimated. Importantly, the calculation of $B\to \pi\pi$ 
amplitudes takes place in full QCD at finite $m_b$ 
and one uses the same input as in the sum rule 
for the $B\to \pi$ form factor. The 
latter provides also the factorizable part 
of the $B\to \pi\pi $ amplitude.

Penguin topology contributions to $B\to \pi\pi$ 
have been estimated using LCSR 
already in previous papers \cite{KMU,KMMpeng},
while annihilation topologies are notoriously difficult 
to calculate. 
This is due to the fact that the factorizable
contribution of the current-current operators 
$O_{1,2}$ with annihilation topology vanishes.
Hence, in 
a sum rule calculation, the leading nonfactorizable 
annihilation with one hard-gluon exchange already corresponds 
to a set of two-loop diagrams with several scales ($b$ quark
mass and external momenta). 

In the present paper we employ a simplified method for the 
annihilation via hard gluons, that is, both final state 
pions are replaced by DA's and only the initial state $B$ meson is 
interpolated by an appropriate current. 
Importantly, the propagators 
appearing in the resulting Feynman diagrams carry 
sufficient virtuality to be treated 
perturbatively. Although this is a modification of the 
original idea presented in \cite{AKBpipi}, it is a 
useful approach, since 
it yields a result free of infrared divergences. 
This result 
may be compared to what is obtained from QCDF, where an infrared 
regulator needs to be introduced, 
reducing basically the whole annihilation contribution to a 
nonperturbative parameter of order $1/m_b$.
On the other hand, the  annihilation with soft-gluon exchange 
which is not accessible in QCDF, is calculated
here within the standard procedure  \cite{AKBpipi} 
leading also to a finite answer. 
After adding the hard and soft-gluon contributions 
from LCSR, we obtain the main result: the finite matrix element
of the operator $O_1$ with annihilation topology, including its phase.
We predict this effect to be numerically small.

Furthermore, we investigate the annihilation contributions 
via quark-penguin operators.
While $O_{3,4}$ have the same $V-A$ structure as $O_{1,2}$,
so that the factorizable annihilation vanishes,
the operators $O_{5,6}$ contribute
through two different types of contractions. One of them has 
a $V+A$ content and also vanishes in the factorizable annihilation, 
whereas the other one, with a $S\pm P$ structure allows for a
factorizable $B\to\pi\pi$ transition with annihilation topology.
This contribution reduces to a separate 
nonperturbative object, the pion scalar form factor at 
timelike momentum transfer $m_B^2$. 
In QCDF and PQCD, the factorizable annihilation  
was taken into account only with a  
perturbative gluon exchange between the final state quarks,
corresponding to the pion form factor in $O(\alpha_s)$. 
The method of LCSR allows to obtain the zeroth order 
in $\alpha_s$, that is the ``soft'' (end-point) 
part of the scalar pion form factor. We calculate this part, 
with the same approach as for the e.m. (vector) 
pion form factor in \cite{ffpion,BK}. The resulting 
hadronic matrix element is large, 
due to the chirally enhanced factor. Still this effect 
alone cannot produce a large \(\Delta I=1/2\)
amplitude in $B\to\pi\pi$, because of the small Wilson 
coefficients. 

Finally, we evaluate the $B\to\pi\pi$ decay amplitudes
using LCSR predictions and including 
all calculated nonfactorizable effects with the emission, 
penguin  and annihilation topology. Since these effects 
are generally small, the discrepancy between 
the $B^0\to\pi^+\pi^-,~\pi^0\pi^0$ 
observables calculated in the factorization limit 
and the current experimental data remains. 
One may encounter a situation similar to $K\to \pi\pi$, 
with its long-standing problem of $\Delta I =1/2$ rule. 

The paper is organized as follows: 
In section \ref{Sec-Pheno-fact} 
we summarize the current status of phenomenology of $B\to \pi\pi$ 
amplitudes, starting from the isospin decomposition and
naive factorization and comparing them with the data.
In section \ref{Sec-Pheno-nonfact} 
we discuss nonfactorizable effects, representing 
the $B\to \pi\pi$ decay amplitudes 
in terms of hadronic matrix elements
of effective operators with different topologies. 
In section \ref{Sec-Method}, we 
derive the LCSR for the hadronic matrix
element of an effective operator with a given topology.
In sections 
\ref{Sec-Hard} and \ref{Sec-Soft} we present
our new results for \(B \to \pi\pi\) annihilation 
with perturbative (hard) and nonperturbative (soft) 
gluons, respectively.  
In section \ref{Sec-O6} we obtain the sum rule
for the pion scalar form factor which determines the 
specific factorizable annihilation contribution of the
operators $O_{5,6}$.
In section \ref{Sec-Numerics}, we perform the numerical 
analysis and present the  
LCSR prediction for the annihilation contributions
to \(\bar{B}^0\to\pi^+\pi^-\).
Furthermore, we add 
all calculated contributions and present our 
numerical predictions for the branching ratios and direct 
CP-asymmetries in 
all three \(B\to\pi\pi\) channels.
In section \ref{Sec-Comparison} we  analyze our 
result in the limit \(m_b\to\infty\) and comment 
on the annihilation mechanism in 
QCDF and PQCD. We conclude in section  \ref{Sec-Concl}.
The appendices contain some expressions used in the paper.

\section{ Phenomenology of $B\to \pi\pi$ amplitudes}
\label{Sec-Pheno-fact}
Throughout this paper we adopt isospin symmetry.
We also neglect the effects of the electroweak penguin
operators, so that the
effective weak Hamiltonian for $B\to \pi\pi$ 
has the following expression:
\ba
H_{\textrm{eff}}=\frac{G_F}{\sqrt{2}}\Big\{
\lambda_u\left(c_1\,O_1^u+c_2\,O_2^u\right)+
\lambda_c\left(c_1\,O_1^c+c_2\,O_2^c\right)\nonumber\\
+
(\lambda_u+\lambda_c)\Big[\sum\limits_{i=3}^{6}c_iO_i +c_{8g}O_{8g}
\Big]\Big\}+ h.c.~~~.
\label{eq-Heff}
\ea
The CKM factors are defined as $\lambda_p= V_{pb}V_{pd}^*$ ($p=u,c,t)$, and
we use the CKM unitarity replacing $-\lambda_t$ by $\lambda_u+\lambda_c$.
Hereafter we suppress for brevity 
the scale dependence in the Wilson coefficients $c_i$,   
which is supposed to be compensated by the scale-dependence
of the hadronic matrix elements of the effective operators $O_i$.
The current-current operators entering Eq.~(\ref{eq-Heff}) 
are 
\ba
O_1^{p}=(\bar{d}\Gamma_\mu p)(\bar{p}\Gamma^\mu b)=\frac{1}3O_2^p+
2\widetilde{O}_2^p\,,~~
O_2^p=(\bar{p}\Gamma_\mu p)(\bar{d}\Gamma^\mu b)=\frac{1}3O_1^p+
2\widetilde{O}_1^p\,,
\label{eq-O12}
\ea
where $p=u,c$, $\Gamma_\mu=\gamma_\mu(1-\gamma_5)$ and 
\ba
\widetilde{O}_1^p=(\bar{d}\Gamma_\mu \frac{\lambda^a}2
p)(\bar{p}\Gamma^\mu\frac{\lambda^a}2 b)\,, ~~
\widetilde{O}_2^p=(\bar{p}\Gamma_\mu \frac{\lambda^a}2
p)(\bar{d}\Gamma^\mu\frac{\lambda^a}2 b)\, 
\label{eq-otilde}
\ea
with $Tr(\lambda_a\lambda_b)= 2\delta_{ab}$.
The color Fierz transformation allows 
to use, instead of the combination $c_1O_1^p+c_2O_2^p$, either
 $(c_1+c_2/3)O_1^p+2c_2\widetilde{O}_1^p$ or  the opposite one
with $1\leftrightarrow 2$. In leading order
the operators with color-neutral currents factorize
and nonfactorizable contributions start from two-gluon exchanges,
which we will neglect. The color-octet currents yield 
nonfactorizable effects starting at a one-gluon level.
These effects will be systematically taken into account.

To complete the definition of $H_{\textrm{eff}}$, 
we specify the quark penguin operators:
\ba
O_3=\sum\limits_f
(\bar{f}\Gamma_\mu f)(\bar{d}\Gamma^\mu b)=\frac{1}3O_4+
2\widetilde{O}_4\,,~~
O_4=\sum\limits_f(\bar{d}\Gamma_\mu f)(\bar{f}\Gamma^\mu b)=
\frac{1}3O_3+2\widetilde{O}_3\,,\nonumber\\
O_5=\sum\limits_f
(\bar{f}\gamma_\mu(1+\gamma_5)f)(\bar{d}\Gamma^\mu b)=\frac{1}3O_6+
2\widetilde{O}_6\,,
\nonumber\\
O_6=-2 \sum\limits_f(\bar{d}(1+\gamma_5)f)(\bar{f}(1-\gamma_5) b)=
\frac{1}3O_5+2\widetilde{O}_5\,,
\label{eq-O36}
\ea
where $f=u,d,s,c,b$ and we again use
the color Fierz decompositions introducing the 
operators with color-octet currents 
$\widetilde{O}_{3,4,5,6}$ obtained 
from $O_{3,4,5,6}$ , respectively.
Finally, the chromomagnetic quark-gluon penguin operator is:
\be
O_{8g}= -\frac{g_s}{8\pi^2}m_b\bar{d}\sigma_{\mu\nu}(1+\gamma_5)
G^{\mu\nu}b\,.
\ee

Turning to the phenomenology of $B\to \pi\pi$ 
we begin with quoting the results 
of the current measurements
\cite{HFAG}:
\begin{eqnarray}\
BR(B^+ \to \pi^+ \pi^0)&=& (5.5 \pm 0.6)\! \times \!10^{-6}\nonumber \\ 
BR(B^0 \to \pi^+ \pi^-)&=& (5.0 \pm 0.4)\! \times \!10^{-6}\nonumber\\
BR(B^0 \to \pi^0 \pi^0) &=& (1.45 \pm 0.29)\! \times\!
10^{-6}\,,
\label{eq-Bpipidat}
\end{eqnarray}
where only the first branching ratio is compatible with 
the expectations of factorization.

The problem may be analyzed in terms of the usual 
isospin decomposition. 
The effective Hamiltonian of the Standard Model consists of two
parts with $\Delta I = 1/2$ and $\Delta I = 3/2$, resulting in two
reduced isospin amplitudes $A_0$ and $A_2$ in \(B\to\pi\pi\),
which correspond
to the pions in the $I=0$ and $I = 2$ final states, respectively.
One obtains the following decomposition
for the amplitudes%
\footnote{
\,Throughout this paper we consider, for definiteness,
$B^-$ and $\bar{B}^0$ decay amplitudes, whereas all quoted
branching ratios for $B^+$ and $B^0$ are, as usual, $CP$-averaged.}
\begin{equation}
\begin{array}{rcl}
A(B^- \to \pi^- \pi^0)=&\langle \pi^-  \pi^0 | H_{\textrm{eff}} | B^-  \rangle
&\displaystyle=\frac{3}{\sqrt{2}}\,  A_2 \, , 
\\[.2cm] 
A(\bar{B}^0 \to \pi^+ \pi^-)=&\langle \pi^+ \pi^-| H_{\textrm{eff}} |
\bar{B}^0 \rangle &\displaystyle=A_2+A_0\,,
\\[.2cm]
A(\bar{B}^0 \to \pi^0 \pi^0)=&\langle \pi^0 \pi^0| H_{\textrm{eff}} |
\bar{B}^0 \rangle &\displaystyle=2A_2-A_0\,, 
\end{array}
\label{eq-isospin}
\end{equation}
from which the well known isospin 
relation \cite{Gronauetal} is obtained:
\be
A(\bar{B}^0 \to \pi^0 \pi^0)=\sqrt{2}A(B^- \to \pi^- \pi^0)-A(\bar{B}^0 \to
\pi^+ \pi^-)\,.
\label{rel}
\ee
In the above we use the same convention for the amplitudes, as in
Ref.~\cite{BBNS1,BBNS2}, including the statistical factor 1/2 in the branching
ratio for $\bar{B}^0 \to \pi^0 \pi^0$.

From Eq.~(\ref{eq-isospin})
one obtains the ratio of  the moduli 
of $A_0 / A_2$ in terms of the decay rates
\begin{equation}
\frac{|A_0|}{|A_2|} = \sqrt{3 
\Big[\frac{BR(\bar{B}^0 \to \pi^0 \pi^0)+BR(\bar{B}^0 \to \pi^+ \pi^-)}{
BR(B^- \to \pi^- \pi^0)}\Big]\frac{\tau_{B^-}}{\tau_{B^0}}-2}\,. 
\label{eq-brratio}
\end{equation}
Using Eq.~(\ref{eq-Bpipidat}) 
(neglecting $CP$ asymmetries)  
and $\tau_{B^-}/\tau_{B^0}= 1.075 \pm 0.009$ 
\cite{HFAG}, one gets 
\begin{equation}
\frac{|A_0|}{|A_2|} =   1.33\pm 0.31\,.
\label{eq-a0a2numb}
\end{equation}

Employing the tree-level emission graphs and retaining 
only the current-current operators 
(the {\em naive factorization} limit)  we may obtain 
a first insight into the anatomy of these decays. In fact, 
this simplified way leads practically 
to the same conclusions as the full calculation in 
the framework of QCDF.
First of all, the decay $B^- \to \pi^- \pi^0$ 
is well described in the factorization limit, where
\be 
\sqrt{2}A(B^- \to \pi^- \pi^0)= \lambda_u\,
\frac43\Big(c_1(\mu)+c_2(\mu)\Big)\,{\cal A}_{\pi\pi}\, \\
\label{eq-c1c2Bpipi}
\ee
with the usual notations for the Wilson coefficients $c_{1,2}$
and the factorizable $B\to \pi\pi$ amplitude 
\be
{\cal A}_{\pi\pi}=i\frac{G_F}{\sqrt{2}}f_\pi
f^0_{B\pi}(m_\pi^2)(m_B^2-m_\pi^2) \,. 
\label{Afact}
\ee
In the above, $f_\pi=131$ MeV is the pion decay constant and  
$f^0_{B\pi}(q^2)$ 
is the scalar $B\to \pi$ form factor. 
Hereafter, we neglect $m_\pi$ in the amplitudes, retaining it only in the
ratio $\mu_\pi=m_\pi^2/(m_u+m_d)$. 

In the factorization approximation, it is sufficient 
to use the leading-order (LO) values of the Wilson
coefficients. We vary their renormalization scale 
within $m_b/2<\mu <m_b$; for illustrative purpose we also 
put $\mu=M_W$. Using the LCSR prediction \cite{Bpi} 
for the $B\to \pi$ form factor: 
\be
f^0_{B\pi}(m_\pi^2)\simeq f^0_{B\pi}(0)=f^+_{B\pi}(0)= 
0.26\pm 0.05\,,
\ee
which is explained below in sect. \ref{Sec-Numerics}, 
and taking $|V_{ub}|=(4.22\pm 0.11 \pm 0.24)\times 10 ^{-3}$ 
from Ref. \cite{CKMfitter} (adding the errors in quadrature)
we obtain from Eq.~(\ref{eq-c1c2Bpipi})
\begin{equation}
BR(B^+ \to \pi^+ \pi^0)_{fact}= 
\left\{ \begin{array}{l} 
 (5.7^{+2.4}_{-2.0}\pm 0.7 ) \times 10^{-6}\,,\,\mu = m_b/2\,\, (c_1=1.169, c_2=-0.361)\\[0.2cm]                       
(6.4^{+2.7}_{-2.3}\pm 0.8) \times 10^{-6}\,,\,\mu = m_b \,\,(c_1=1.108, c_2=-0.249)\\[0.2cm]
(8.7^{+3.7}_{-3.0}\pm 1.1) \times 10^{-6}\,, \,\mu = M_W\,\, (c_1=1,c_2=0)
  \, ,  \end{array} \right.  
\end{equation}
where the errors reflect the uncertainties of the 
form factor and of $|V_{ub}|$, respectively. 
The scale-dependence is mild, as expected.

Another way to check the validity of the factorization approximation 
for this channel (independent of $|V_{ub}|$ and the value of $f^+_{B\pi}(0)$)
is provided by the ratio of the $B^+ \to \pi^+ \pi^0$
and $B^0 \to \pi^- l^+ \nu_l$ widths. In factorization
approximation:
\ba
\frac{BR(B^+ \to \pi^+ \pi^0)}{BR(B^0\to \pi^-l^+\nu_l)}
=\frac{2\pi^2|V_{ud}|^2\Big(c_1(\mu)+c_2(\mu)\Big)^2 
f_\pi^2m_B^3}{3\!\int_{\raisebox{-1mm}{$\scriptstyle
0$}}^{\raisebox{0.7mm}{$\scriptstyle(m_B-m_\pi)^2$}}
dq^2\,(E_\pi^2-m_\pi^2)^{3/2}|N_{B\pi}(q^2)|^2}
\Big(\frac{
\tau_{B^+}}{\tau_{B_0}}\Big)\,,
\ea
where $E_\pi=(m_B^2+m_\pi^2-q^2)/(2m_B)$ and 
$N_{B\pi}(q^2)= f^+_{B\pi}(q^2)/f^+_{B\pi}(0)$
is the shape of the form factor.
The recent measurement \cite{BaBarSL} of the
$B\to \pi l \nu$  decay distribution,   
fitted to the parameterization \cite{BecKaid},
$N_{B\pi}(q^2)= [(1-q^2/m_{B^*}^2)(1-\alpha_{B\pi}q^2/m_{B^*}^2)]^{-1}$,
yields $\alpha_{B\pi}=0.61\pm 0.09$. Using this value,  
and the average experimental number $BR(B^0\to \pi^-l^+\nu_l)=
(1.36\pm 0.11)\times 10^{-4}$ \cite{HFAG}, we obtain 
at $  \mu=m_b/2~[m_b;M_W]$:
\ba
BR(B^+ \to \pi^+ \pi^0)_{fact}&=&(3.6 \pm 0.3\pm 0.4)\times 10^{-6}
\,\nonumber\\ 
&&[(4.1 \pm 0.3\pm 0.5)\times 10^{-6};(5.6 \pm 0.5\pm 0.6)\times
10^{-6}]\,
,\label{eq-BRfactSL}
\ea
where the first error originates from the semileptonic branching ratio,
and the second one from the slope parameter $\alpha_{B\pi}$. 
The result  is again in the ballpark of the experimental interval
(\ref{eq-Bpipidat}). 
Hence we conclude that the amplitude $A_2$ 
may be estimated correctly by naive factorization. 

Furthermore, using naive factorization, we may 
express the ratio  $A_0 / A_2$ in terms of the Wilson coefficients, since  all
hadronic matrix elements will drop out. 
Using Eq.~(\ref{eq-c1c2Bpipi}) and the analogous relation 
\be
A(\bar{B}^0 \to \pi^+ \pi^-)=\lambda_u
\Big(c_1(\mu)+\frac{c_2(\mu)}3\Big){\cal A_{\pi\pi}}\,,
\label{eq-c1c2Bpipi1}
\ee
and comparing with the decomposition (\ref{eq-isospin}) we obtain 
\begin{equation}
\frac{A_0}{A_2} = \frac{5}{4} \Bigg(\frac{c_1(\mu) - c_2(\mu)/5}{c_1(\mu) +
c_2(\mu)}\Bigg)= \left\{ \begin{array}{l} 
 1.92 \,,\mu = m_b/2\\                       
1.68\,,\mu = m_b \\
 1.25\,, \mu = M_W\, .  \end{array} \right.  
\end{equation}
In fact, this expression depends quite strongly on the scale, 
showing that naive factorization for $B^0$ modes 
is compatible with the data only
for a large scale of $O(m_W)$, which seems unrealistic.

A more distinct disagreement 
is revealed between the ratios of $B^0$ 
and $B^+$ partial widths 
calculated in the naive factorization at 
$\mu=m_b/2~[m_b;M_W]$:
\ba
\frac{BR(B^0\to \pi^+\pi^-)}{BR(B^+ \to \pi^+ \pi^0)}
=1.77~[1.49; 1.05]\,,~~
\frac{BR(B^0\to \pi^0 \pi^0)}{BR(B^+ \to \pi^+ \pi^0)}
\simeq 10^{-3}~[0.010; 0.058]\,,
\ea 
and the same ratios obtained from the experimental results
(\ref{eq-Bpipidat}). In fact, the calculated 
$B\to \pi^0\pi^0$ width is too small even at $\mu=m_W$. 
We conclude that the naive factorization picture misses
an important part of the amplitude 
$A_0$ which 
interferes destructively 
(constructively) with $A_2$ in $A(\bar{B}^0\to \pi^+\pi^-)$ 
($A(\bar{B}^0\to \pi^0 \pi^0)$). 
If there were large nonfactorizable gluon corrections 
to the emission topology, they would have influenced both $A_0$ and 
$A_2$, violating the abovementioned agreement for 
the $B^-\to \pi^-\pi^0$ channel. 
Hence, the missing isospin zero amplitude
should be searched for  within the contributions of 
non-emission topologies for the current-current 
operators and/or in the contributions of the penguin operators, 
an opinion shared by many recent analyses of these decays.

\section{Beyond Factorization}
\label{Sec-Pheno-nonfact}

\begin{figure}[htb]
\begin{center}
\subfigure[emission]{\includegraphics[scale=0.5]{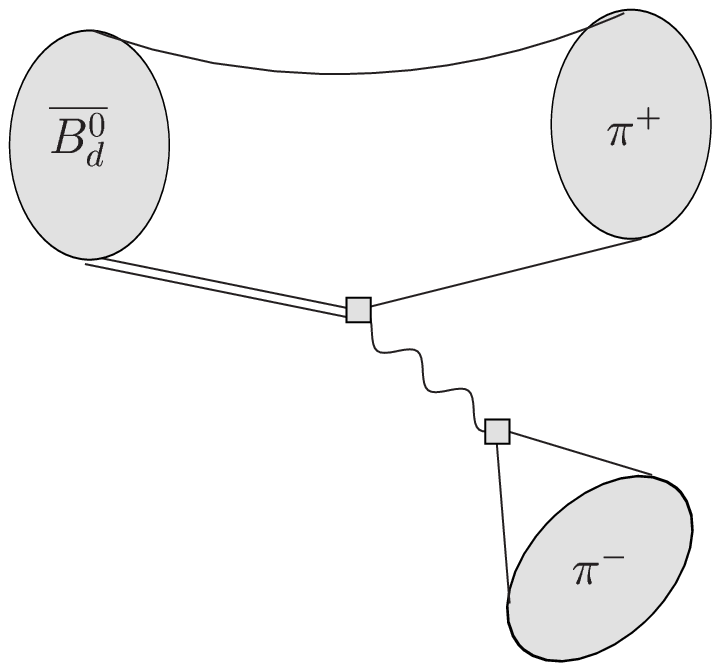}}
\hspace{0.5cm}
\subfigure[penguin]{\includegraphics[scale=0.5]{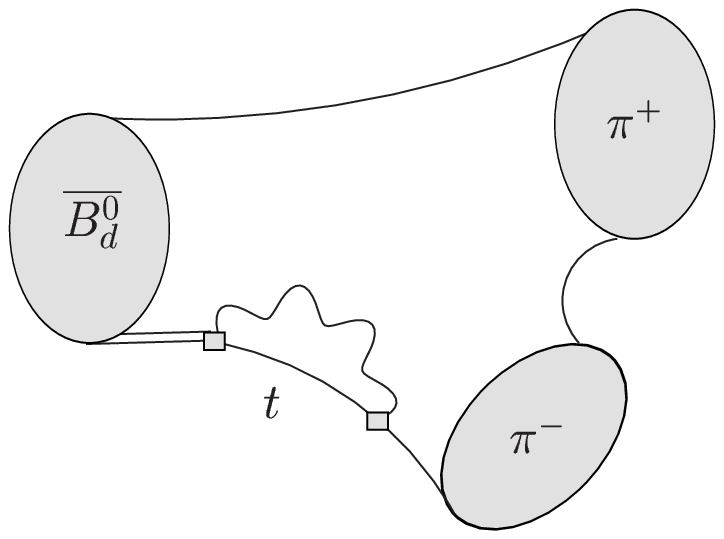}}
\subfigure[charming
penguin]{\hspace{0.5cm}\includegraphics[scale=0.5]{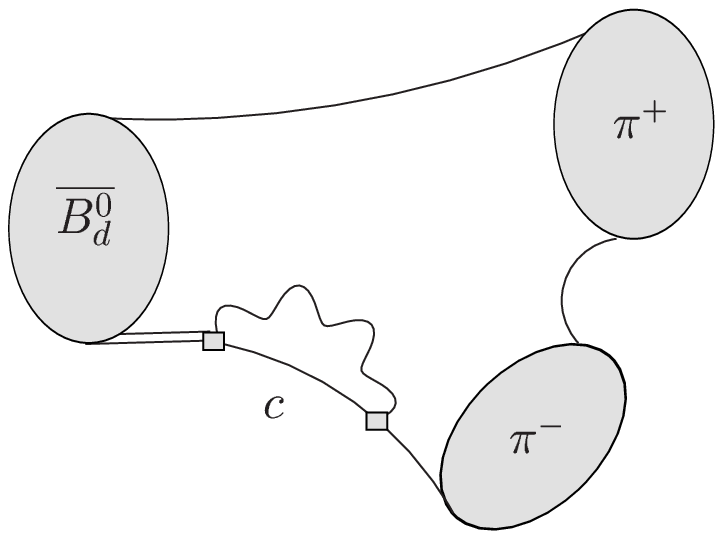}}\\
\subfigure[annihilation]{\includegraphics[scale=0.5]{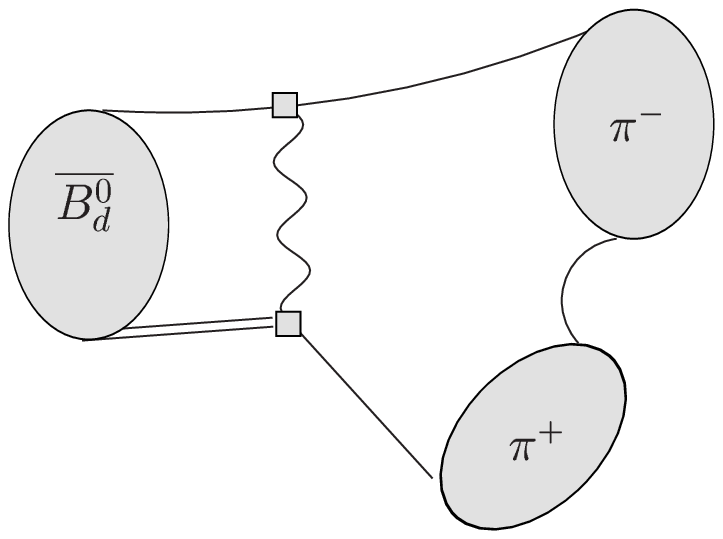}}
\subfigure[penguin
annihilation]{\hspace{0.5cm}\includegraphics[scale=0.5]{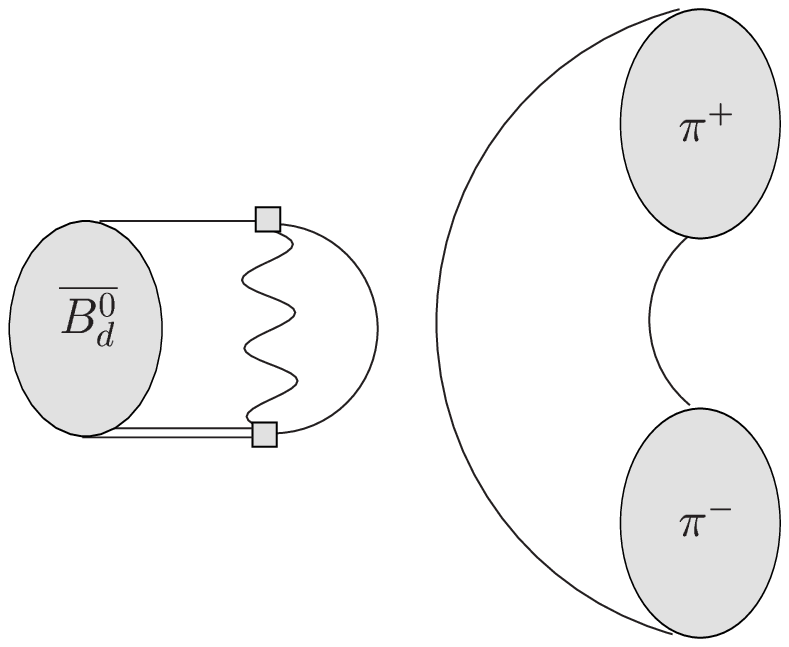}}
\end{center}
\caption{\em Different quark topologies in \(\bar{B}^0\to\pi^+\pi^-\); 
double lines denote the $b$ quark, wavy lines the W-boson. }
\label{fig-Bpipitop}
\end{figure}

In Fig.~\ref{fig-Bpipitop} we schematically represent
different quark topologies contributing to the amplitude 
of $\bar{B}^0\to \pi^+\pi^-$, the channel of our interest.
Under {\em topology} we understand the way to contract the  
valence quarks (antiquarks) of the initial $B$ and final mesons
with the antiquarks (quarks) from the effective 
weak Hamiltonian. This concept is meaningful for all those
methods where the valence quark structure of mesons 
is well defined, either via the meson distribution amplitudes 
(DA's) as, e.g. in QCDF,  or via both DA's and interpolating currents,
as in the correlation function for LCSR. 
Denoting the valence spectator antiquark in $\bar B$ by 
$\bar{q_s}$ and the quarks emitted in the b-quark decay 
by $q_1,q_2,\bar{q_3}$ we define
{\em emission } as the part of the decay amplitude 
where all four quarks and antiquarks end up as the valence
quarks of the final mesons (Fig.~\ref{fig-Bpipitop}a). 
Correspondingly {\em penguin} is the 
part where $\bar{q_s}$  and only one of $q_1,q_2$ belong 
to the final mesons (Figs.~\ref{fig-Bpipitop}b,c).
The remaining two possibilities are {\em annihilation}
(Fig.~\ref{fig-Bpipitop}d) 
or {\em penguin annihilation } (Fig.~\ref{fig-Bpipitop}e),
where either $\bar{q_3}$ and the one of $q_1,q_2$  
or none of the quarks are among the valence
quarks of the final mesons, respectively.
\footnote{\,Note that drawing generic quark-line diagrams, 
one can always start 
from the emission topology and then, merging the quark-antiquark 
(spectator) lines in the final state, end up with 
an annihilation mechanism; according to our classification, 
this mechanism  still belongs to annihilation.}

In the isospin symmetry limit it is sufficient 
to investigate in detail the $\bar{B}^0\to \pi^+\pi^-$ amplitude.
It has a single $I=2$ , $O(\lambda_u)$ part with emission topology,   
common with the $B^-\to \pi^-\pi^0$ amplitude, and in addition 
contains many different $I=0$ contributions proportional to 
both $\lambda_u$ and $\lambda_c$. The $\bar{B}^0\to \pi^0\pi^0$ 
amplitude is then obtained by simply using the isospin 
relation (\ref{rel}).

The complete isospin decomposition of the $\bar{B}^0 \to \pi^+ \pi^-$
amplitude following from Eq.~(\ref{eq-Heff}) can be cast
in the following form:
\ba
A(\bar{B}^0 \to \pi^+ \pi^-)= \lambda_u \Big( A^{(u,1,2)}_2+ 
A^{(u,1,2)}_0\Big) +\lambda_c A^{(c,1,2)}_0
+(\lambda_u+\lambda_c) A^{(\geq 3)}_0\,, 
\label{eq-decomptot}
\ea
where the upper indices in $A_I^{(i)}$ indicate the contributing 
operators and the lower
index the isospin. Correspondingly,
\be 
\sqrt{2}A(B^-\to \pi^-\pi^0)=\lambda_u (3 A^{(u,1,2)}_2).
\label{eq-decomp1}
\ee
Each separate amplitude $A_I^{(i)}$ 
in Eqs.~(\ref{eq-decomptot},\ref{eq-decomp1})
contains a sum over hadronic matrix elements 
with different topologies ($T$): emission ($E$), 
penguin ($P_q$,$P_c$,$P_b$), 
annihilation ($A$) and penguin annihilation ($P_qA$,$P_cA$,$P_bA$),
where $q=u,d,s$. 
According to Ref.~\cite{BurasSilvestr}, this 
subdivision can be 
systematically done in a scheme- and scale-independent way.
It is clear, that the number of independent matrix elements 
of the type $\langle \pi^+ \pi^- |O_i |\bar{B}^0 \rangle_T $  
is less than the number of operators because certain penguin 
operators have the same quark structure as the 
current-current operators. For the penguin topologies we will also 
neglect the differences between quark loops with $q=u,d,s$. 

Let us first discuss the \(I=2\) part
in Eqs.~(\ref{eq-decomptot}) and (\ref{eq-decomp1}), 
which is relatively simple. 
Taking into account the nonfactorizable emission correction,
where 
only $\tilde{O}_1^u$ 
contributes in the one-gluon approximation, we introduce the ratio:
\be
r_E^{(\pi\pi)}=\frac{\langle\pi^+\pi^-|\tilde{O}_1^u
|\bar{B}^0 \rangle_E }{\langle \pi^+\pi^-| O_1^u|\bar{B}^0 \rangle_E }\,.
\label{eq-rEdef}
\ee
Note that in the adopted approximation
the matrix element standing in the denominator is factorizable:  
$G_F/\sqrt{2}\langle\pi^+\pi^-| O_1^u| \bar{B}^0\rangle_E ={\cal
A}_{\pi\pi}$.
We obtain:
\be 
 A^{(u,1,2)}_2= 
\Big[ \frac49\Big(c_1+c_2\Big)+\frac23\Big(c_1+c_2\Big)r_E^{(\pi\pi)}\Big]
{\cal A}_{\pi\pi}\,.
\label{eq-iso2Bpipi}
\ee
Accordingly, the $I=0$  part generated by the operators $O_{1,2}^u$ 
has the following decomposition:
\ba
A^{(u,1,2)}_0=\Big[
\frac19\Big(5c_1-c_2\Big)-\frac23\Big(c_1-2c_2\Big)r_E^{(\pi\pi)}
+2c_1\Big(r_{P_u}^{(\pi\pi)}+r_A^{(\pi\pi)}\Big)\Big]{\cal A}_{\pi\pi}\,,
\label{eq-iso0Bpipi}
\ea
where the relative contributions for penguin and annihilation 
topologies \footnote{\,In this paper, we neglect the penguin
annihilation (PA) topologies. They can be
simply added to the general decomposition by introducing 
the corresponding  $r_{PA}^{(\pi\pi)}$ ratios, 
but from LCSR we expect them to 
be small; this mechanism is also neglected in QCDF. 
} 
similar to Eq.~(\ref{eq-rEdef})
are defined as follows:
\be
r_{P_u}^{(\pi\pi)}=\frac{\langle\pi^+\pi^-|\tilde{O}_2^u
|\bar{B}^0 \rangle_{P_u} }{\langle \pi^+\pi^-| O_1^u|\bar{B}^0 \rangle_E }\,,
~~
r_A^{(\pi\pi)}=\frac{\langle\pi^+\pi^-|\tilde{O}_2^u
|\bar{B}^0 \rangle_A }{\langle \pi^+\pi^-| O_1^u|\bar{B}^0 \rangle_E }\,.
\label{eq-rArP}
\ee
Hereafter, we will use 
a more generic notation $r_{P_q}^{(\pi\pi)}$, $q=u,d,s$,
so that $r_{P_u}^{(\pi\pi)}=r_{P_q}^{(\pi\pi)}$
in accordance with our approximation for the light-quark
penguin loops. 
Furthermore, the $O_{1,2}^c$ 
operators with $c$-quark contribute 
to Eq.~(\ref{eq-decomptot}) only in the penguin topology
(``charming penguins''):
\be
A^{(c,1,2)}_0= 2c_1r_{P_c}^{(\pi\pi)}\,,
~~\mbox{where}~~ 
r_{P_c}^{(\pi\pi)}=\frac{\langle\pi^+\pi^-|\tilde{O}_2^c
|\bar{B}^0 \rangle_{P_c} }{\langle \pi^+\pi^-| O_1^u|\bar{B}^0 \rangle_E }\,.
\label{eq-rPc}
\ee

The remaining piece of the decomposition
(\ref{eq-decomptot}) containing the
hadronic matrix elements of quark-penguin operators
with various topologies, normalized to the factorizable part, 
is more complicated:
\begin{multline}
A^{(\geq 3)}_0=\Bigg[ c_4+\frac{c_3}3
+2c_3\Big(r_E^{(\pi\pi)}+ r_{P_q}^{(\pi\pi)}+
r_{P_b}^{(\pi\pi)}+r_A^{(\pi\pi)}\Big)
+2c_4\Big( 3\bar{r}_{P_q}^{(\pi\pi)}+\bar{r}_{P_c}^{(\pi\pi)}+
\bar{r}_{P_b}^{(\pi\pi)}+2r_A^{(\pi\pi)}\Big)
\\
+\frac{2\mu_\pi}{m_b}\Big(c_6+\frac{c_5}3\Big)+
2c_5r_E^{(\pi\pi,6)}
+ 2c_6\Big(3\bar{r}_{P_q}^{(\pi\pi)}+\bar{r}_{P_c}^{(\pi\pi)}+
\bar{r}_{P_b}^{(\pi\pi)}+
2r_A^{(\pi\pi,5)}\Big)
\\
+\Big(c_6+\frac{c_5}3\Big)R_A^{(\pi\pi,6)}+ 2c_5r_A^{(\pi\pi,6)}
+c_{8g}^{eff}r_{8g}^{(\pi\pi)}\Bigg]{\cal A}_{\pi\pi}\,.
\label{eq-3456Bpipi}
\end{multline}
Some of the $r^{(\pi\pi)}_T$ -parameters 
entering the above equation have already been defined:
we use  the fact that certain quark-penguin and 
current-current operators coincide.
The parameter $r_{P_b}^{(\pi\pi)}$ determines 
the relative contribution of the $b$-quark penguin topology. 
It is defined as in Eq.~(\ref{eq-rPc}) with $c\to b$. 
Furthermore, the notation 
$\bar{r}_{P_{q,c,b}}^{(\pi\pi)}$
is introduced to distinguish the 
penguin contractions of the operators
$O_{4,6}$ ($\tilde{O}_{3,5}$) from 
from those of $O_{1,3}$ ($\tilde{O}_{2,4}$). 
In the NDR scheme used here, as in \cite{BBNS1,BBNS2},
the quark loop factors for the two contractions 
differ by a constant.
The loop factor entering LCSR for $r_{P_{q,c,b}}^{(\pi\pi)}$ 
is given in Eq.~(4) of \cite{KMMpeng}. To obtain the 
corresponding factor for $\bar{r}_{P_{q,c,b}}^{(\pi\pi)}$ 
one simply has to subtract 1/6 from this expression.
A few terms
in Eq.~(\ref{eq-3456Bpipi}) are generated by the 
effective operator $O_6$ with $(S+P)\otimes(S-P)$ structure. 
First, the factorizable emission 
contribution of the $u$-quark part 
of this operator, $O_6^u=-2 (\bar{d}(1+\gamma_5)u)(\bar{u}(1-\gamma_5)
b)$, 
acquires a ``chirally-enhanced'' factor 
$\mu_\pi/m_b$, whereas 
its nonfactorizable part is described by an 
additional parameter:
\be
r_E^{(\pi\pi,6)}=\frac{\langle\pi^+\pi^-|\tilde{O}_6^u
|\bar{B}^0 \rangle_E }{\langle \pi^+\pi^-| O_1^u|\bar{B}^0 \rangle_E }\,.
\label{eq-rEdef1}
\ee
Second, there is a specific factorizable annihilation 
contribution due to the $d$-quark part of the same operator 
$O_6^d\equiv-2 (\bar{d}(1+\gamma_5)d)(\bar{d}(1-\gamma_5) b)$ 
which is expressed in terms of the 
$B$ meson decay constant 
$\langle 0 |\bar{d}\gamma_5 b |B \rangle=-im_B^2f_B/m_b$ 
multiplied by the pion scalar form factor: 
\be   
\langle \pi^+(p_1)\pi^-(p_2) \mid \bar{d}d \mid 0 \rangle=
F_\pi^S((p_1+p_2)^2)\,, 
\label{scalarFF}
\ee
corresponding to the transition of the scalar
and isoscalar quark-antiquark current into a two-pion state 
with the invariant mass squared $(p_1+p_2)^2=m_B^2$.
The parameter in Eq.~(\ref{eq-3456Bpipi})
determining the annihilation via $O_6^d$ is 
factorized as:
\be
R_A^{(\pi\pi,6)}=\frac{\langle\pi^+\pi^-|O_6^d
|\bar{B}^0 \rangle_A }{\langle \pi^+\pi^-| O_1^u|\bar{B}^0 \rangle_E }=
-\frac{2f_B F_\pi^S(m_B^2)}{m_b f_\pi f_{B\pi}^+(0)}\,.
\label{matrO6}
\ee
The nonfactorizable annihilation correction of the color-octet counterpart
of this operator in Eq.~(\ref{eq-3456Bpipi}) is parameterized as
\be
r_A^{(\pi\pi,6)}=\frac{\langle\pi^+\pi^-|\tilde{O}_6^d
|\bar{B}^0 \rangle_A }{\langle \pi^+\pi^-| O_1^u|\bar{B}^0 \rangle_E }\,.
\label{eq-rArP6}
\ee
In addition to this \(S\pm P\) contractions, the nonfactorizable annihilation
contribution from the \(V+A\) contraction is defined in analogy to
Eq.~(\ref{eq-rArP6}) with (\(6\to 5\)).
Finally, the parameter in  Eq.~(\ref{eq-3456Bpipi})
describing the contribution of the gluonic
penguin operator (with the penguin topology) is:
\be
r_{8g}^{(\pi\pi)}=\frac{\langle\pi^+\pi^-|O_{8g}
|\bar{B}^0 \rangle_{P_{g}}}{\langle \pi^+\pi^-| O_1^u|\bar{B}^0 \rangle_E }\,.
\label{eq-r8g}
\ee
In the NDR scheme it should be multiplied by 
$c_{8g}^{eff}=c_{8g}-c_5$ (see \cite{BBNS2}).

For convenience, in Appendix A we present the 
relations of the parameters $r_T^{(\pi\pi)}$
introduced above to the effective coefficients 
$a_i$ and $b_i$ used 
in QCDF \cite{BBNS1,BBNS2} to encode the nonfactorizable effects 
in $B\to \pi\pi$. Eq.(\ref{eq-decomptot}) can also be converted
into  a  typical decomposition in terms of ``tree'', 
``color-suppressed''  and ``penguin'' amplitudes 
used in the CP-analysis of charmless decays,
where the separation to $\lambda_u=|\lambda_u|e^{-i\gamma}$ 
and $\lambda_c$ parts is made explicit:
\be
A(\bar{B}^0 \to \pi^+ \pi^-)= e^{-i\gamma}T_{\pi\pi} + P_{\pi\pi}\,,~~
\sqrt{2}A(B^- \to \pi^- \pi^0)= e^{-i\gamma}(T_{\pi\pi} + C_{\pi\pi})\,.
\label{eq-PCT1}
\ee
Comparing the above with Eq.~(\ref{eq-decomptot})
one reads off:
\ba
T_{\pi\pi}=|\lambda_u|\Big(A^{(u,1,2)}_2+ 
A^{(u,1,2)}_0 +A^{(\geq 3)}_0\Big)\,, 
\nonumber\\ 
P_{\pi\pi}=\lambda_c \Big(A^{(c,1,2)}_0 +A^{(\geq 3)}_0\Big)\,,~~ 
T_{\pi\pi} + C_{\pi\pi}=|\lambda_u|\Big(3A^{(u,1,2)}_2\Big)\,.
\label{eq-PCT2}
\ea

To analyze the $B\to \pi\pi$ amplitudes in terms of 
separate isospin contributions, 
one needs the numerical values of the hadronic parameters entering
Eqs.~(\ref{eq-iso2Bpipi}),(\ref{eq-iso0Bpipi}), (\ref{eq-rPc}) and
(\ref{eq-3456Bpipi}). 
Some of them have already been estimated 
using LCSR. For the nonfactorizable emission entering through
\(r_E^{(\pi\pi)}\)
we will partly use the QCDF result.
The calculation of the unknown annihilation parameters
\(r_A^{(\pi\pi)}\) and \(R_A^{(\pi\pi,6)}\) from LCSR is the main issue of
this paper.

\section{Derivation of LCSR}
\label{Sec-Method}
The method we apply is basically the one developed in 
\cite{AKBpipi}, (see also \cite{KMU,KMMpeng})
with some important modifications which will be 
explained in this section.
To demonstrate the derivation of LCSR for 
a generic hadronic matrix element 
$\langle \pi^+\pi^-|O|\bar{B}^0\rangle_T $,
let us choose for definiteness the 
combination of current-current operators $O=c_1O^u_1+c_2O^u_2$
at a fixed scale, considering it as a superposition of two 
local operators.

One starts from the correlation function
\begin{equation}
  F^{(O)}_\alpha(p,q,k)=-\int d^4x~e^{-i(p-q)x}\int d^4y~e^{i(p-k)y}
  \langle 0|T\left\{j_{\alpha 5}^{(\pi)}(y) O^u(0)
j_5^{(B)}(x)\right\}|\pi^-(q)\rangle,
  \label{eq-F}
\end{equation}
where \(j_{\alpha 5}^{(\pi)}=\bar{u}\gamma_\alpha\gamma_5 d\) and
\(j_5^{(B)}=m_b \bar{b}i\gamma_5 d\)
are the quark currents interpolating the pion and the \(B\) meson,
respectively.
The momentum \(k\) is artificial 
and will vanish in the final sum rule. It
is introduced in order to have 
two independent kinematical variables in the \(B\) and  \(\pi\pi\)
channels.

The correlator (\ref{eq-F})
can be decomposed into four different Lorentz structures,
\[ F_\alpha^{(O)}=(p-k)_\alpha F^{(O)}+
q_\alpha \tilde{F}_1^{(O)}+ k_\alpha \tilde{F}_2^{(O)}
+\epsilon_{\alpha\beta\gamma\rho}q^\beta p^\lambda k^\rho\tilde{F}_3^{(O)},\]
of which we use only the first.
Concerning the kinematical variables, we put \(q^2=m_\pi^2=0\) and choose
\(p^2=k^2=0\) for simplicity.
The remaining invariants are \((p-k)^2, (p-q)^2\) and
\(P^2\equiv(p-k-q)^2\).
In the domain where all three variables are spacelike and large,
all distances are close to the light-cone, \(x^2\sim y^2\sim (x-y)^2\sim 0\),
and the correlation function can be calculated by perturbatively expanding
the \(T\)-product of operators.
In this way, the correlation function is
expressed in a usual form of hard-scattering 
amplitudes convoluted with pion DA's of growing twist.

\begin{figure}[p]
\begin{center}
\subfigure[]{\includegraphics[scale=0.5]{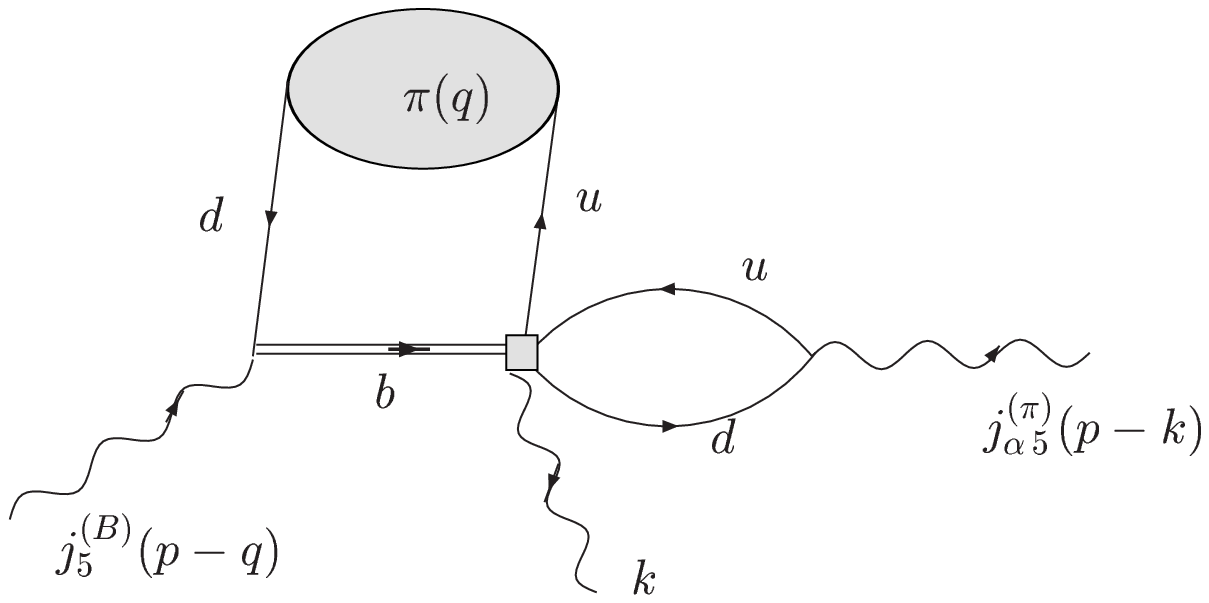}}
\subfigure[]{\hspace{0.5cm}
\includegraphics[scale=0.5]{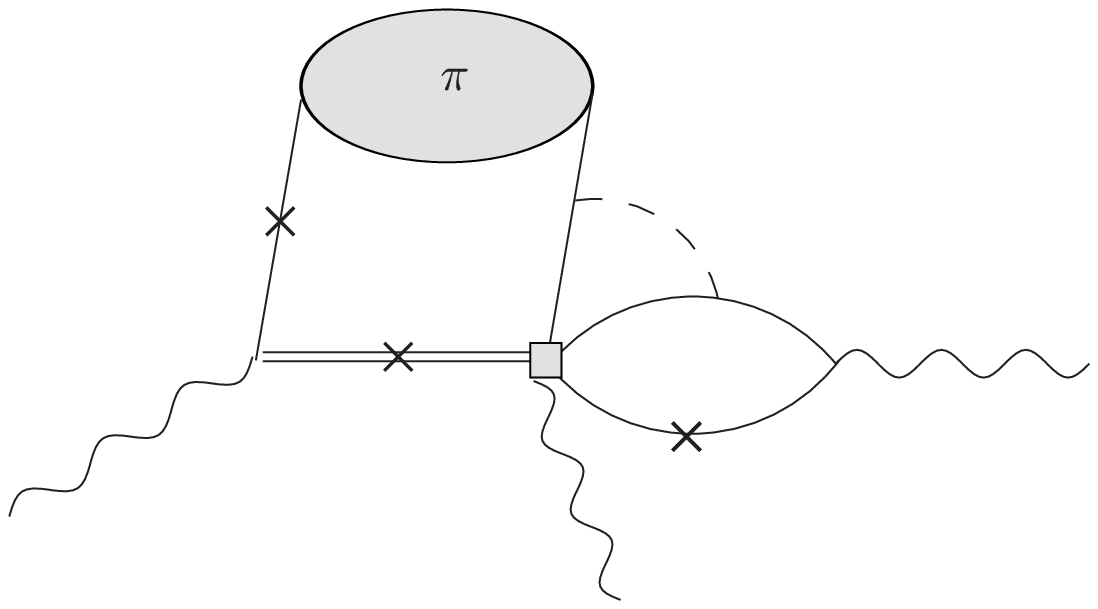}}\\
\subfigure[]{\includegraphics[scale=0.5]{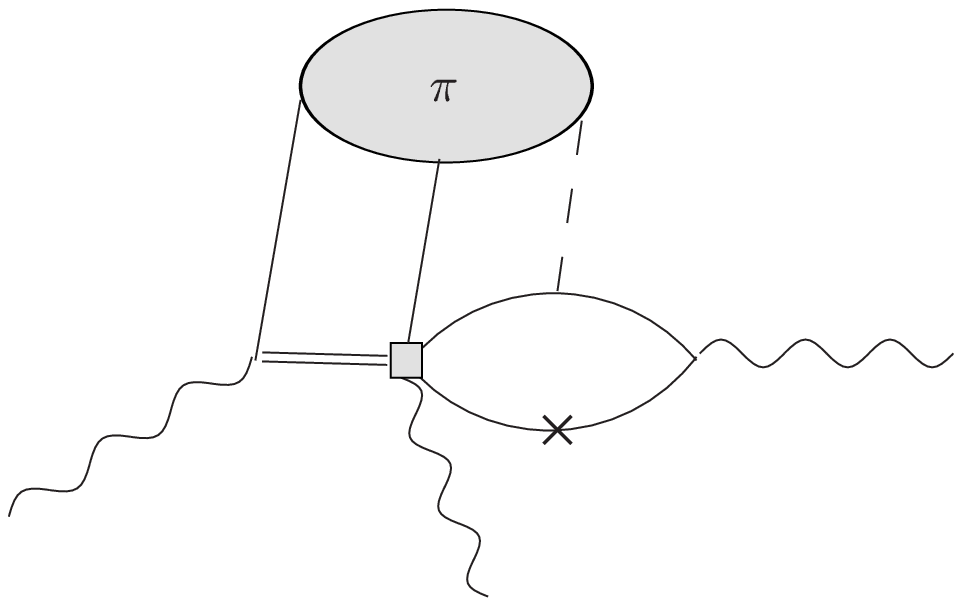}}
\end{center}
\caption{\em 
Diagrams corresponding to the emission topology 
in the OPE of the correlation function (\ref{eq-F}):
(a) factorizable; (b) with nonfactorizable hard gluon 
(six diagrams);(c) nonfactorizable soft gluon 
(two diagrams). The solid, double, dashed, wavy lines
and the square denote the light quarks, b quark, gluon,
external currents and the weak vertex, respectively.
The shaded ovals denote the pion DA's. The crosses
indicate how gluon lines are attached in  the other 
possible diagrams.}
\label{fig-OPE_E}
\end{figure}
\begin{figure}[p]
\begin{center}
\subfigure[]{\includegraphics[scale=0.5]{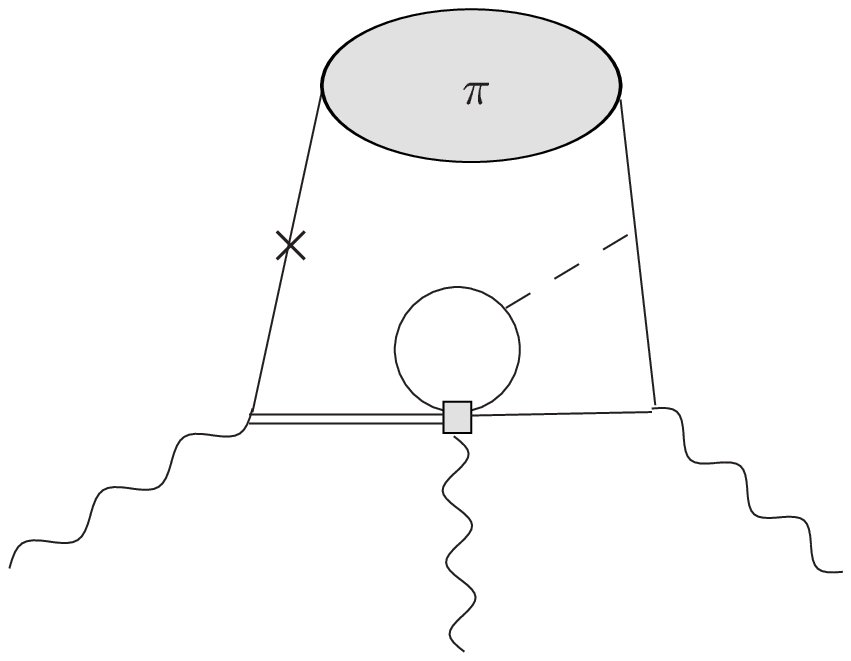}}
\hspace{0.5cm}
\subfigure[]{\includegraphics[scale=0.5]{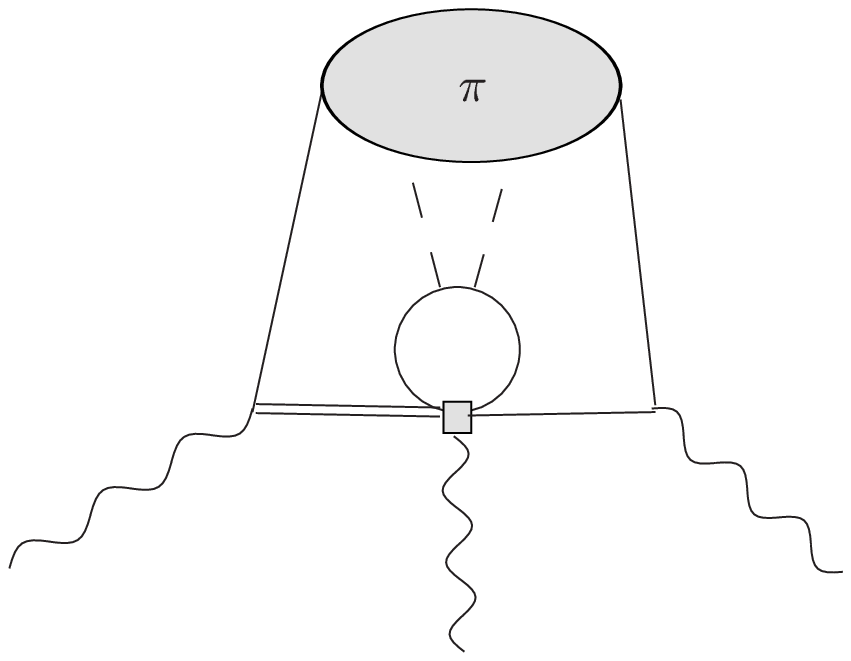}}\\
\end{center}
\caption{\em 
Examples of diagrams corresponding to the 
penguin topology:
with (a) hard gluon and (b) soft gluons.}
\label{fig-OPE_P}
\end{figure}
\begin{figure}[p]
\begin{center}
\subfigure[]{\includegraphics[scale=0.5]{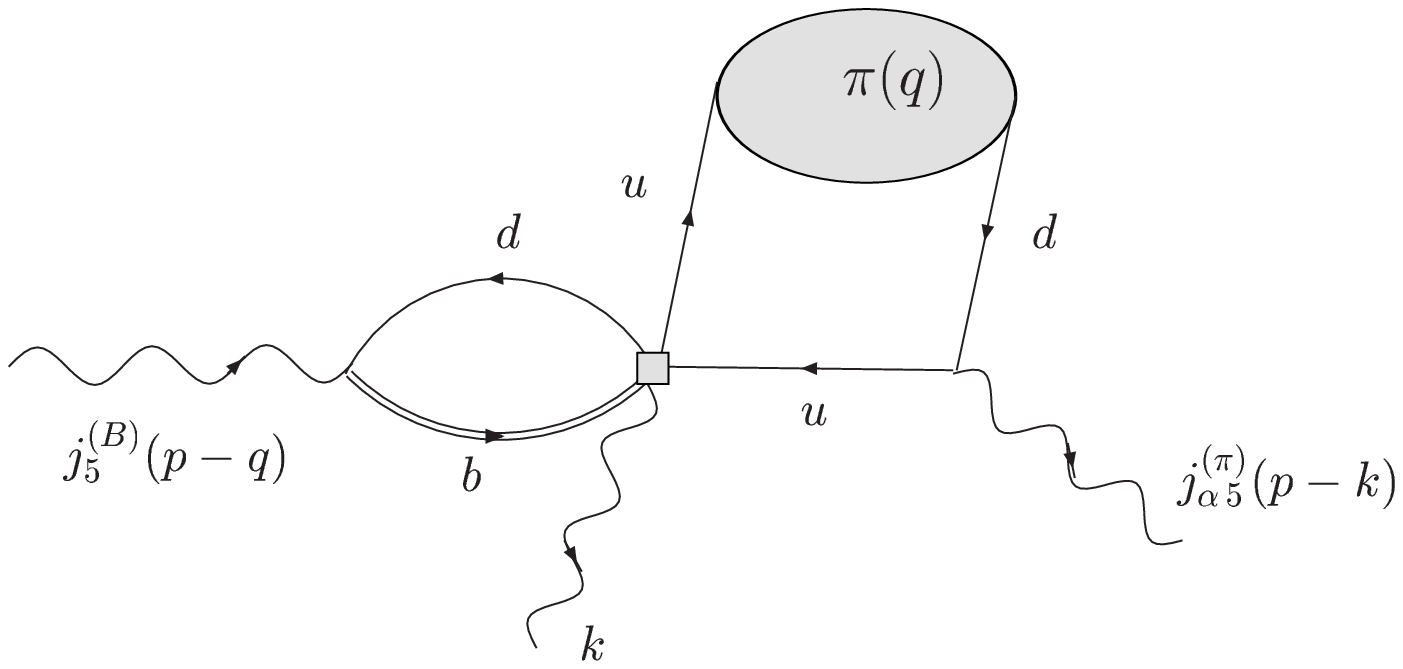}}\\

\subfigure[]{\includegraphics[scale=0.5]{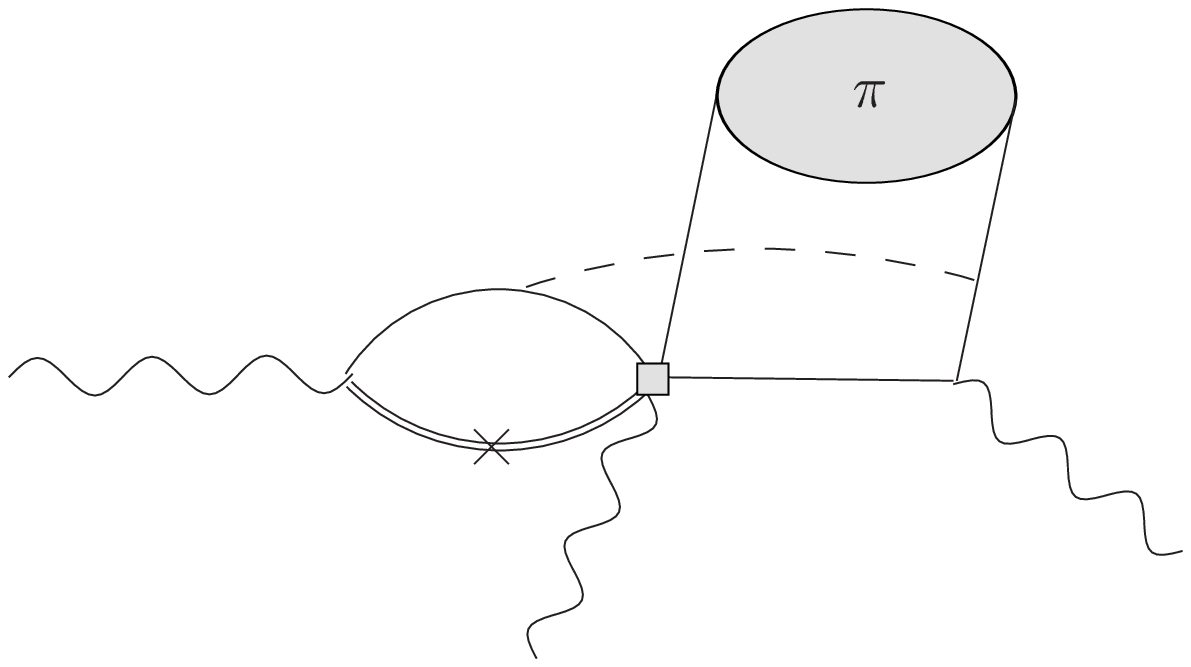}}
\hspace{0.5cm}
\subfigure[]{\includegraphics[scale=0.5]{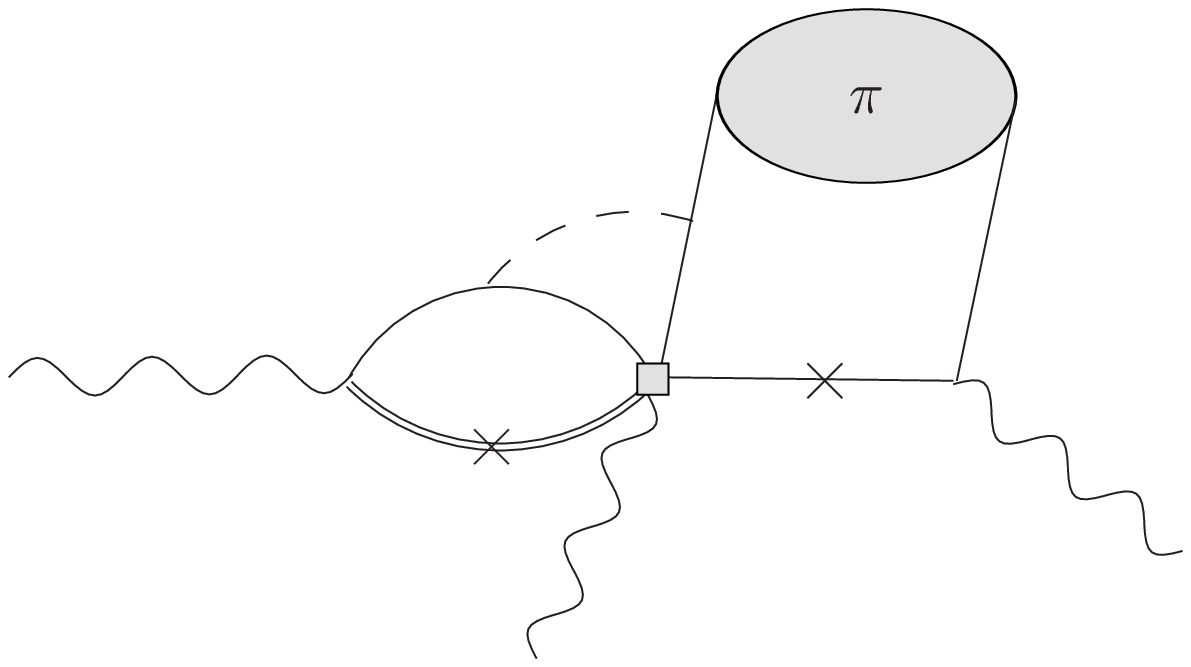}}\\

\subfigure[]{\includegraphics[scale=0.5]{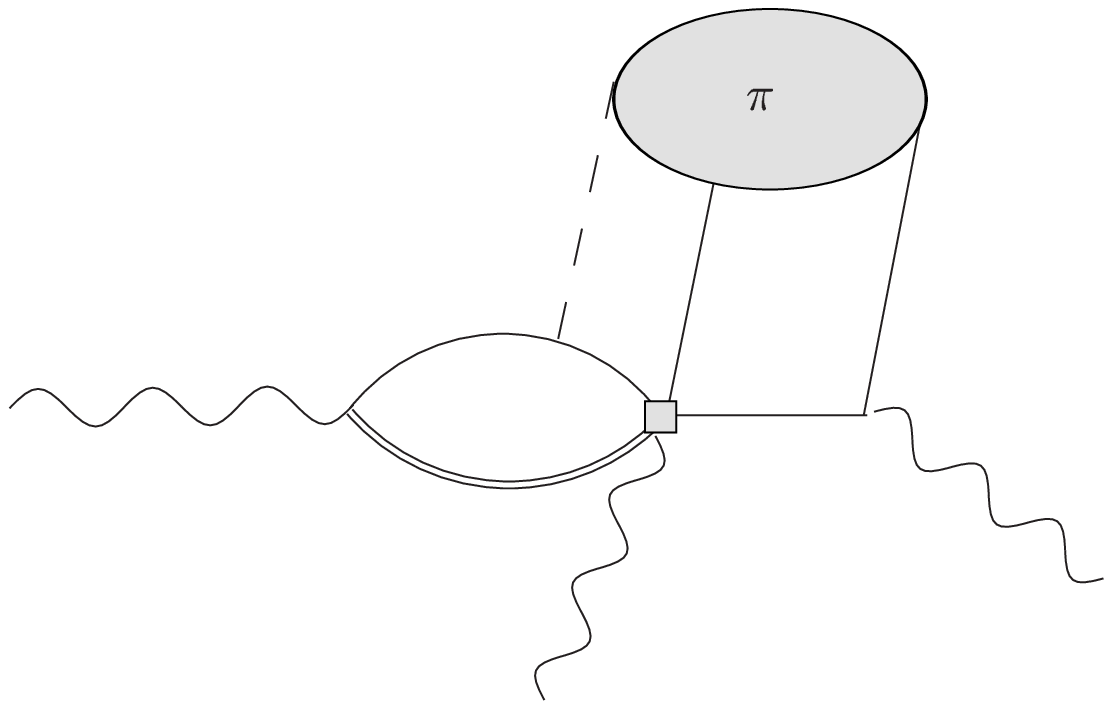}}
\hspace{0.5cm}
\subfigure[]{\includegraphics[scale=0.5]{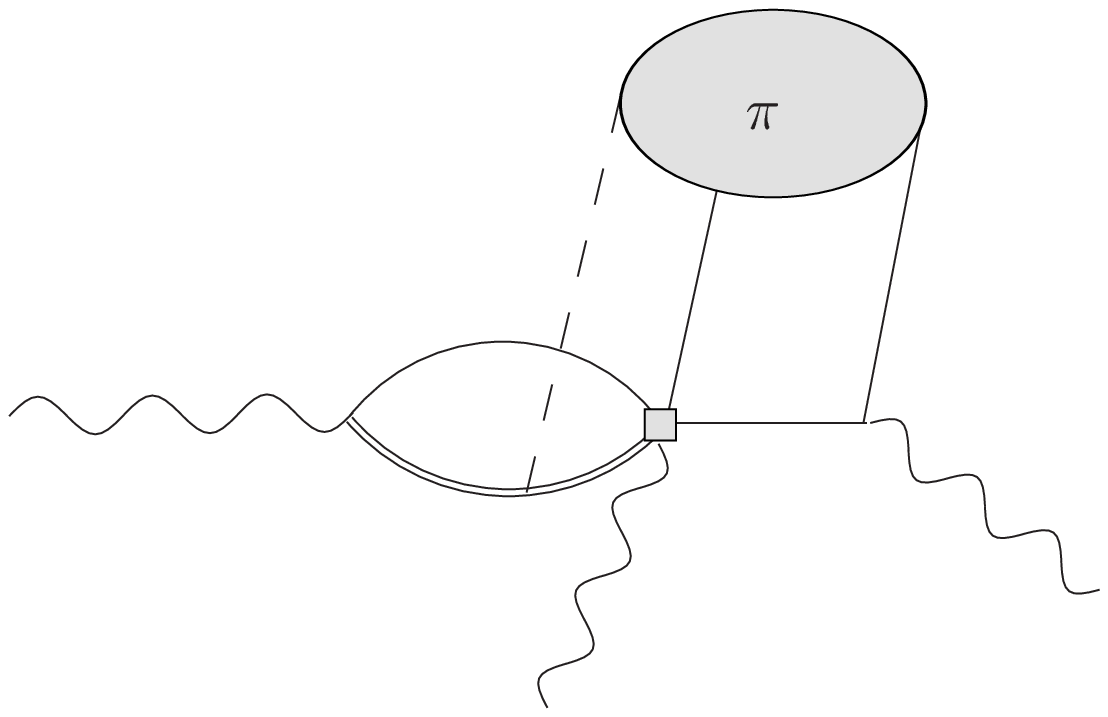}}
\end{center}
\caption{\em 
Diagrams corresponding to the annihilation topology 
in the OPE of the correlation function (\ref{eq-F}):
(a) factorizable; (b),(c) with hard gluon; 
(d),(e) with soft gluon. }
\label{fig-OPE_A}
\end{figure}
\begin{figure}[p]
\begin{center}
\includegraphics[scale=0.5]{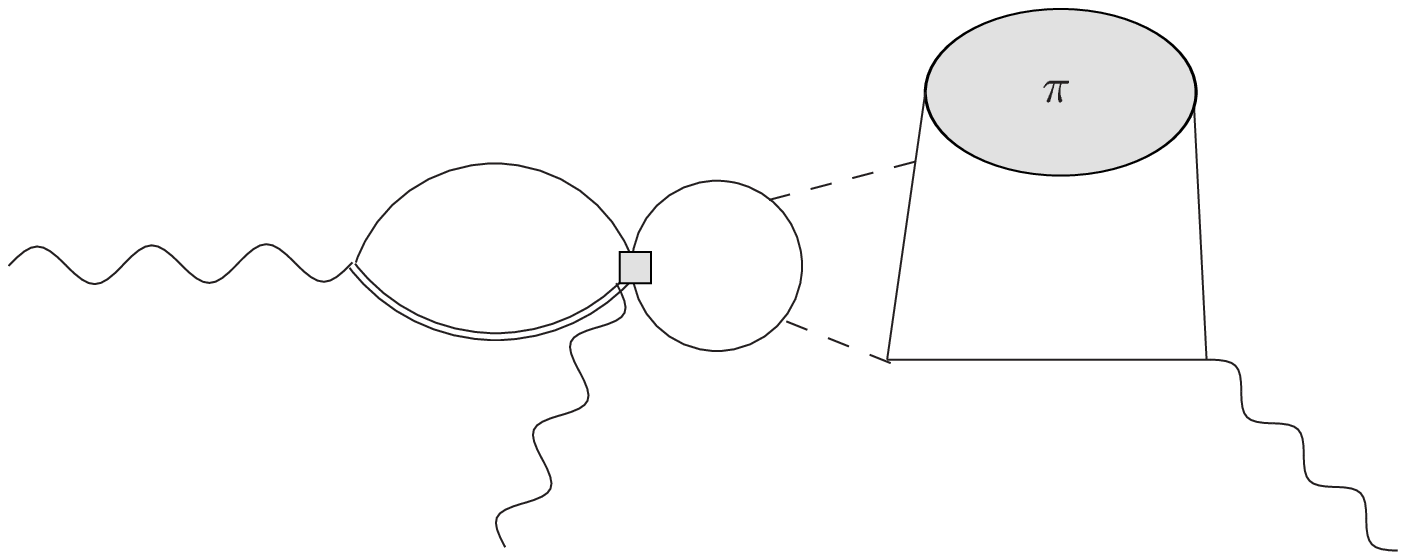}
\hspace{0.5cm}
\includegraphics[scale=0.5]{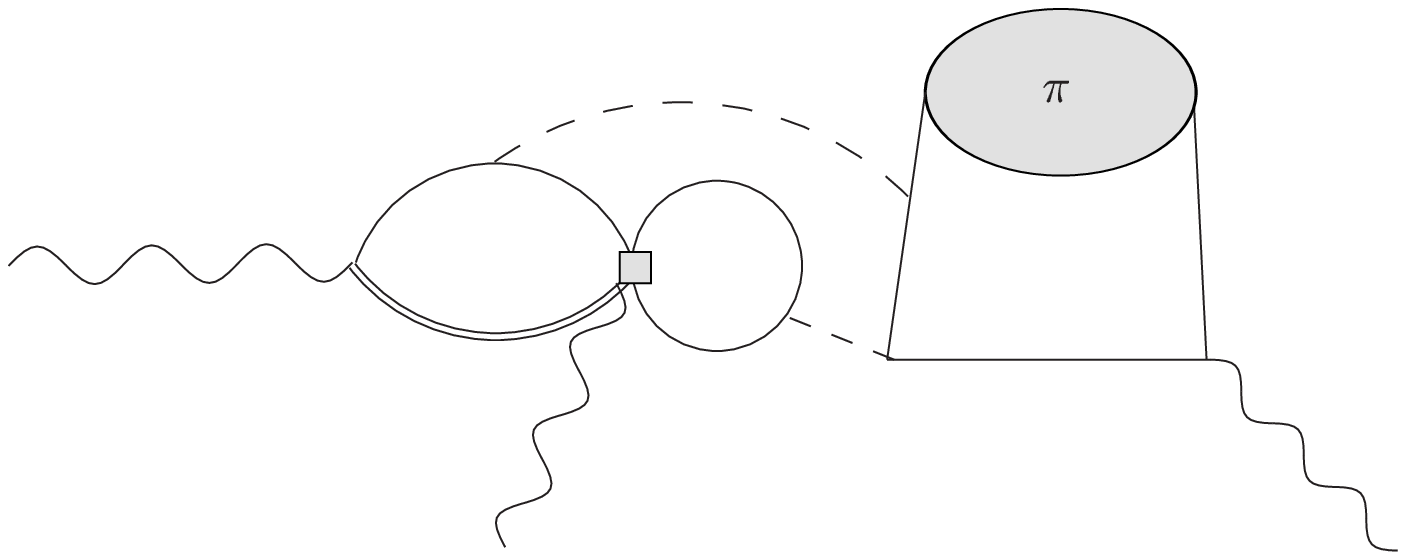}
\end{center}
\caption{\em
Some of the lowest-order diagrams 
corresponding to the 
penguin annihilation topology.}
\label{fig-OPE_PA}
\end{figure}

For a given operator $O$  
in the correlation function (\ref{eq-F}), 
various contractions of the quark fields are possible,
leading to diagrams with different topologies.
Collecting the lowest-order contributions to OPE, 
we easily recognize diagrams with the emission (Fig.~\ref{fig-OPE_E}), 
penguin (Fig.~\ref{fig-OPE_P}),
annihilation (Fig.~\ref{fig-OPE_A}) and penguin-annihilation
(Fig.~\ref{fig-OPE_PA}) topologies. 
The sum of all diagrams calculated at a definite order
in OPE, will be matched to the dispersion relation 
where the ground-state 
contribution contains the hadronic matrix element 
$\langle \pi^+\pi^-|O|\bar{B}^0\rangle $.
If one retains only diagrams with a topology $T$ in the OPE,
the sum rule result  (within adopted accuracy)
can be interpreted
as   $\langle \pi^+\pi^-|O|\bar{B}^0\rangle_T $.%
\footnote{\,
Since we are studying only the leading-order effects here
and use a fixed scale, the
scale- and scheme dependence as well as
the mixing effects between separate operators in $H_{\textrm{eff}}$,
remain beyond our scope. To account for these
effects one has to consider scale- and scheme invariant 
combinations of matrix elements with different topologies
as explained in \cite{BurasSilvestr}.} %
This was actually done for the emission topology 
in \cite{AKBpipi} where the diagrams in  Fig.~\ref{fig-OPE_E}
have been investigated. It was shown that 
retaining only the diagram of Fig.~\ref{fig-OPE_E}a, without gluons
connecting the light-quark loop and the heavy-light part, 
one reproduces the result of naive factorization.
Importantly, the gluons which do not violate factorization 
in this diagram can be added arbitrarily.
Altogether, one obtains the product 
of the LCSR for the $B\to \pi$ form factor 
and the two-point sum rule for 
the pion decay constant. 
The diagrams in Fig.~\ref{fig-OPE_E}b,c describe nonfactorizable corrections
in the emission topology. The diagrams with soft gluons in
Fig.~\ref{fig-OPE_E}c were calculated in \cite{AKBpipi}.
Furthermore, penguin contractions for $O_{1,2}^p$ 
(some of diagrams are shown in Fig.~\ref{fig-OPE_P}) 
have also been studied in the framework of LCSR \cite{KMMpeng}
allowing to calculate the parameters $r^{(\pi\pi)}_{P_c}$,
$r^{(\pi\pi)}_{P_b}$ and $r^{(\pi\pi)}_{P_q}$.
In addition, the LCSR for the gluonic penguin 
operator was derived in \cite{KMU} yielding 
$r^{(\pi\pi)}_{8g}$. 
In the next two sections, we will present the calculation of
the remaining diagrams with the annihilation topology (Fig.~\ref{fig-OPE_A}).

\begin{figure}[bt]
\begin{center}
\subfigure[]{\includegraphics[scale=0.5]{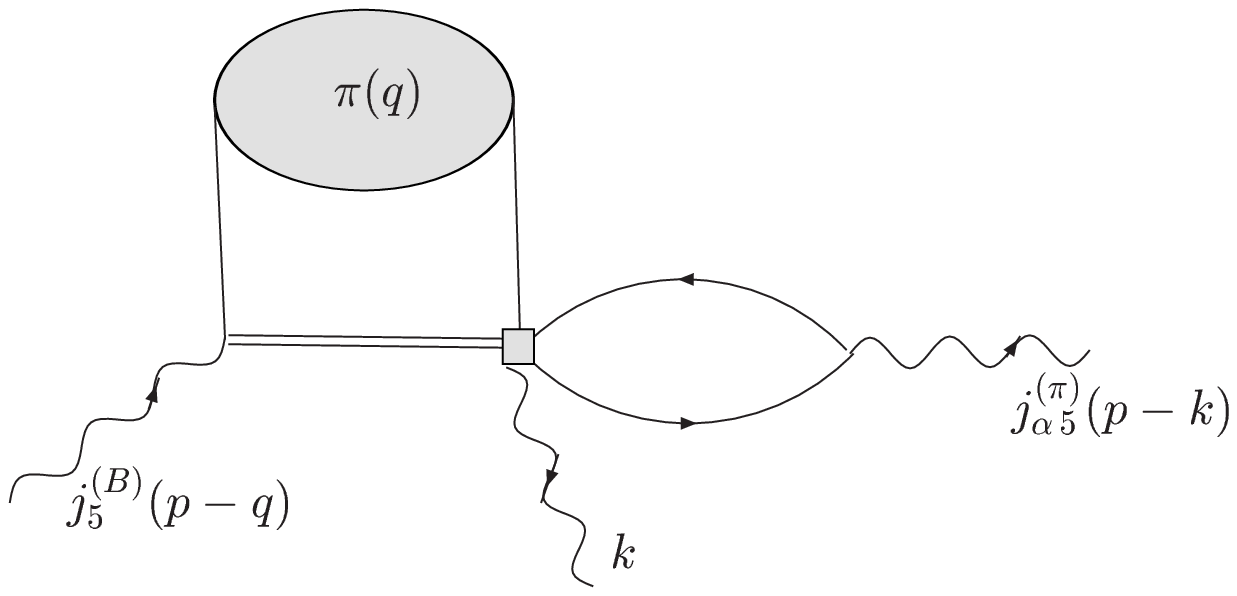}}
\hspace{0.5cm}
\subfigure[]{\includegraphics[scale=0.5]{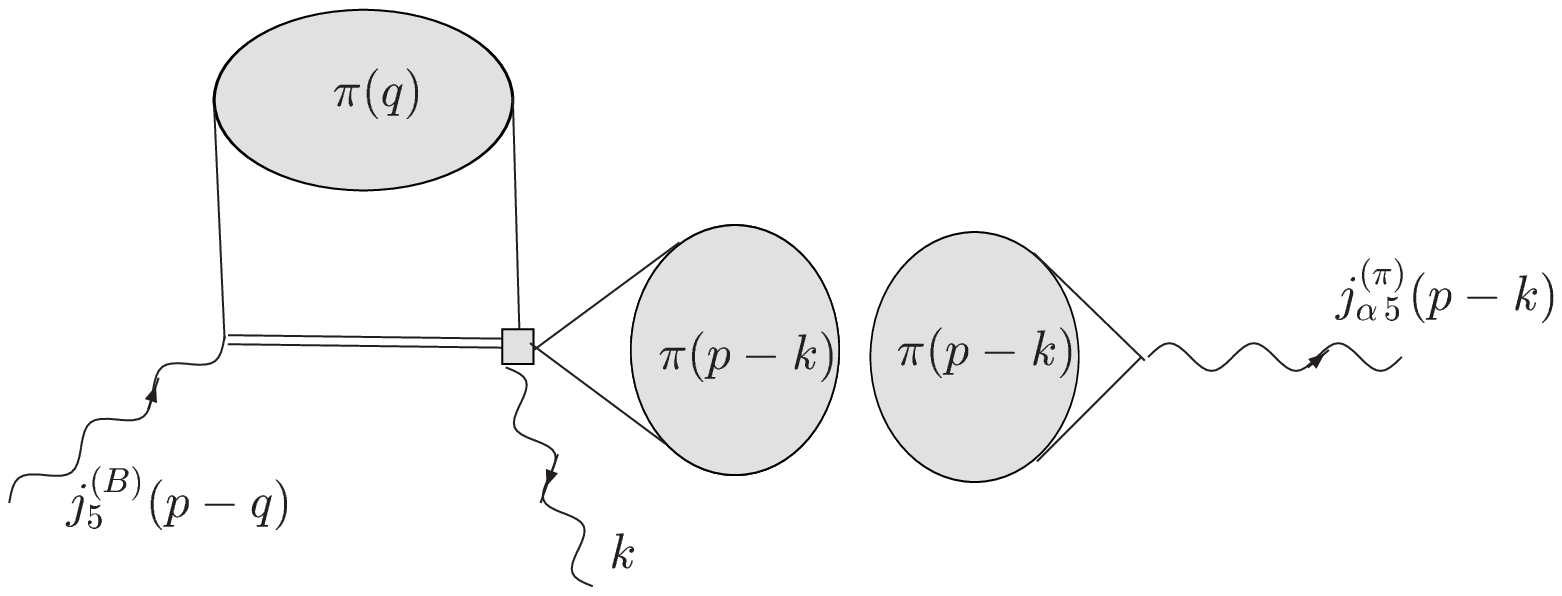}}
\end{center}
\caption{\em The first step in the derivation 
of the sum rule for $B\to \pi\pi$ amplitude in 
the emission topology: (a) the initial correlation function
(only diagrams with emission topology are included) is matched
to (b) the hadronic dispersion relation in the pion channel where 
only the ground-state pion contribution is shown.}
\label{fig-SR1}
\end{figure}

Having at hand the QCD calculation of a set of diagrams
with topology $T$ 
in terms of pion DA's and hard scattering amplitudes, 
one can then express the correlation function 
in the form of a dispersion relation in the variable \(s\equiv (p-k)^2\):
\begin{equation}
  F_{QCD}^{(O,T)}\left((p-k)^2,(p-q)^2,P^2\right)
  =\frac{1}{\pi}\int\limits_0^\infty ds\frac{\textrm{Im}_s
F_{QCD}^{(O,T)}\left(s,(p-q)^2,P^2\right)}
  {s-(p-k)^2-i\epsilon}.
  \label{eq-F-disp}
\end{equation}
On the other hand, one can insert a complete set of hadronic states in the
\(\pi\) meson channel, and obtain
\begin{equation}
  F^{(O,T)}\left((p-k)^2,(p-q)^2,P^2\right)=\frac{i\,f_\pi
\Pi_{\pi\pi}^{(O,T)}\left((p-q)^2,P^2\right)}{-(p-k)^2}
  +\int\limits_{s_h^\pi}^\infty ds\,
\frac{\rho_h^\pi\left(s,(p-q)^2,P^2\right)}{s-(p-k)^2},
  \label{eq-F-had}
\end{equation}
where the one-pion ground state contribution 
contains the pion decay constant and the matrix
element
\begin{equation}
  \Pi_{\pi\pi}^{(O,T)}\left((p-q)^2,P^2\right)
  =i\int d^4x~e^{-i(p-q)x}\langle \pi^-(p-k)|T\left\{O(0)
j_5^{(B)}(x)\right\}|\pi^-(q)\rangle_T\,,
  \label{eq-Pipipi-Corr}
\end{equation}
and $\rho^\pi_h$ is the spectral density of 
heavier hadronic states in this channel. Replacing the integral
over $\rho^\pi_h$ with the standard duality approximation
and equating (\ref{eq-F-disp}) to (\ref{eq-F-had}),
we obtain, after the usual Borel transformation:
\begin{equation}
  \Pi_{\pi\pi}^{(O,T)}\left((p-q)^2,P^2\right)=\frac{-i}{\pi
f_\pi}\int\limits_0^{s_0^\pi}ds~e^{-s/M^2}
  \textrm{Im}_s F_\textrm{QCD}^{(O,T)}\left(s+i\epsilon,(p-q)^2,P^2\right),
  \label{eq-Pipipi-SR}
\end{equation}
where $s_0^\pi$ is the duality threshold in the pion channel.
This first step in the derivation of LCSR is 
schematically shown in Fig.~\ref{fig-SR1} where 
the diagrams with emission topology are chosen for definiteness. 

\begin{figure}[t]
\begin{center}
\subfigure[]{\includegraphics[scale=0.5]{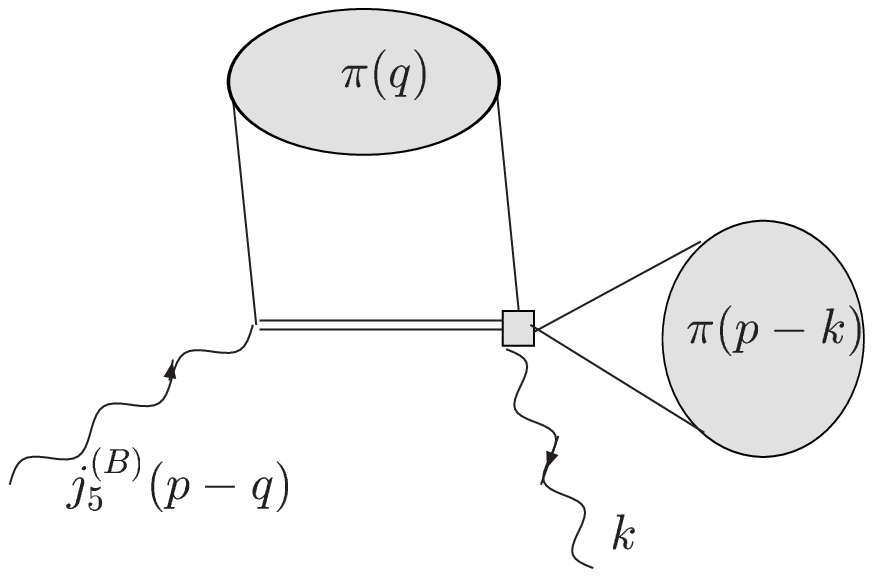}}
\hspace{0.8cm}
\subfigure[]{\includegraphics[scale=0.5]{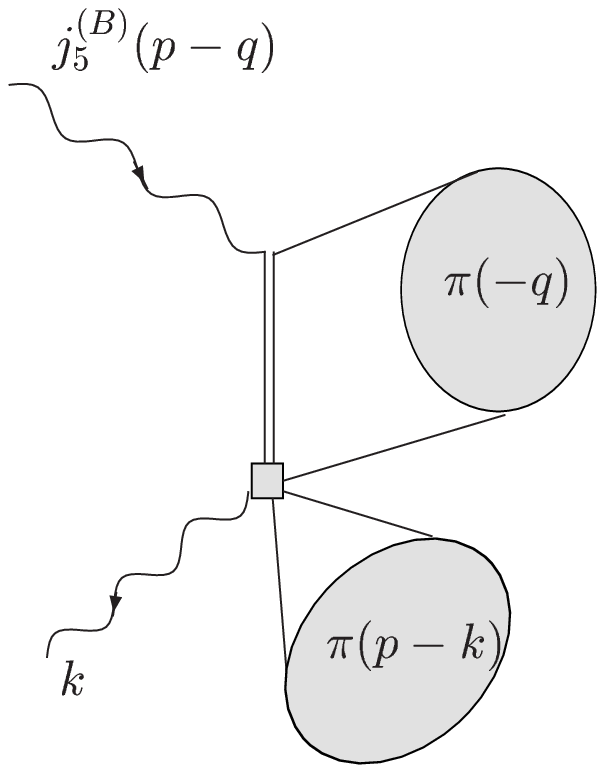}}
\end{center}
\caption{\em The second step in the derivation 
of the sum rule: (a) the two-pion  
matrix element calculated in the spacelike region 
is analytically continued to (b) the same 
matrix element in the timelike region.}
\label{fig-SR2}
\end{figure}

\begin{figure}[t]
\begin{center}
\subfigure[]{\includegraphics[scale=0.5]{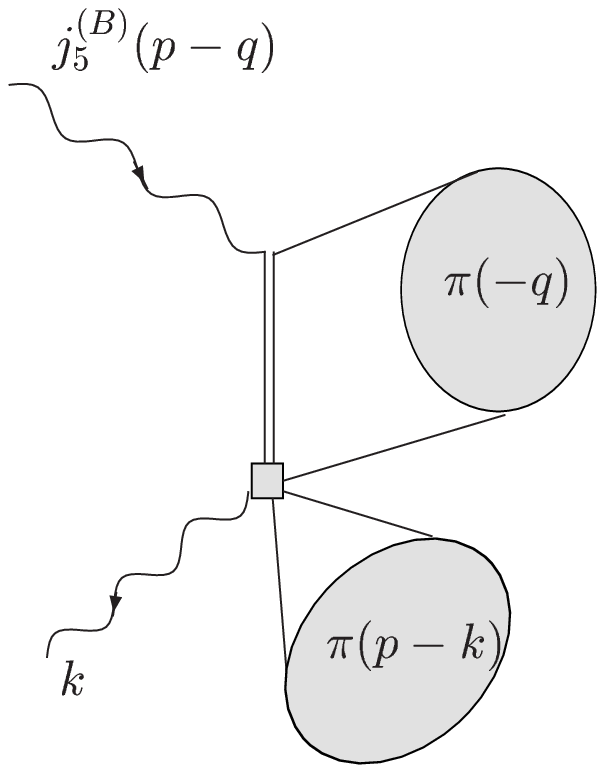}}
\hspace{0.5cm}
\subfigure[]{\includegraphics[scale=0.5]{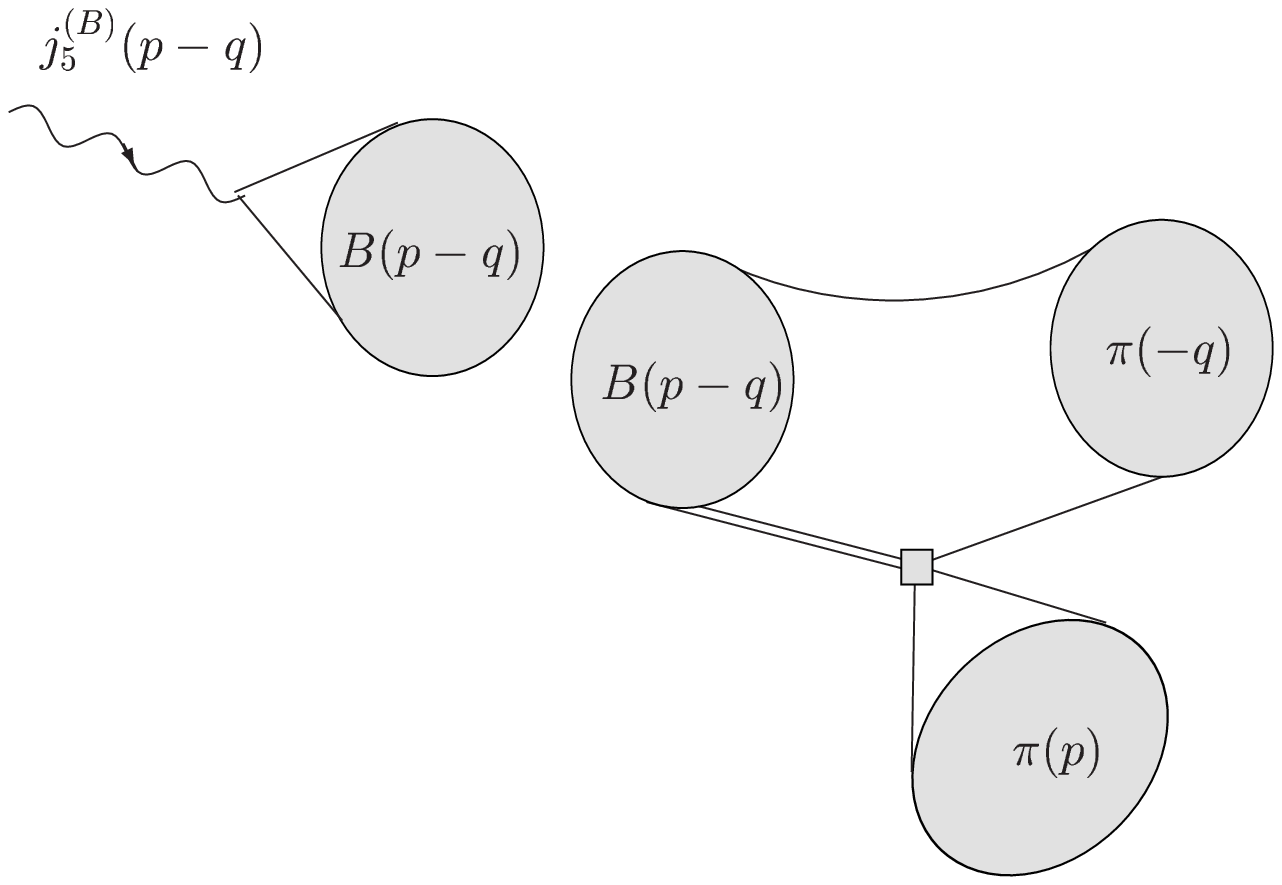}}
\end{center}
\caption{\em The third step in the derivation 
of the sum rule:(a) the two-pion matrix element
$\Pi_{\pi\pi}^{(O,E)}$ 
after analytic continuation to $(p-q-k)^2 =m_B^2$ 
is matched to (b) the hadronic dispersion relation in the 
$B$ meson channel;
the ground state $B$ meson contribution is proportional
to the $B\to \pi\pi$ matrix element in emission topology.}
\label{fig:Bpipitop}
\end{figure}

Note that the hadronic matrix element (\ref{eq-Pipipi-Corr})
diagrammatically shown in Fig.~\ref{fig-SR1}b 
itself represents a correlator. 
At large negative $(p-q)^2$ and $P^2$  it can be 
factorized into a short-distance part (the b-quark propagator) 
and a long-distance part (the combination of two pion DA's).
Adding hard-gluon exchanges does not
seemingly spoil this factorization. In any case,
the absence of infrared singularities 
has to be checked by a direct calculation.  
Hence, instead of using
(\ref{eq-Pipipi-SR}),
we are free to use \(\Pi_{\pi\pi}^{(O,T)}\) as a starting QCD object.
That will be done below in the case of the annihilation topology
with hard gluons, while employing the standard derivation of 
(\ref{eq-Pipipi-SR}) for the soft-gluon part.

To continue, following \cite{AKBpipi}
we consider \(\Pi_{\pi\pi}^{(O,T)}((p-q)^2,P^2)\), 
as an analytical function of the variable \(P^2\),
the invariant mass of the \(\pi\pi\) pair.   
Starting from our calculation for negative \(P^2\), we have to reach 
the physical timelike point \(P^2=m_B^2\) by analytical continuation (see Fig.~\ref{fig-SR2}).
As already explained in \cite{AKBpipi}, we use the fact that
$m_B^2\gg m_\pi^2,s_0^\pi$, i.e., the two-pion system in 
the $B$ decay is in the timelike asymptotic region, far from 
the light-quark resonances. 
Note that at fixed $(p-q)^2$, the matrix element 
\(\Pi_{\pi\pi}^{(O,T)}\) can always be represented in a 
form of a hadronic dispersion relation 
in the variable \(P^2\):
\begin{equation}
  \Pi_{\pi\pi}^{(O,T)}((p-q)^2,P^2)=\frac{1}{\pi}\int\limits_0^\infty dt~
\frac{\textrm{Im}_t
\Pi_{\pi\pi}^{(O,T)}\left((p-q)^2,t\right)}{t-P^2-i\epsilon}\,.
\end{equation}
This dispersion relation is only needed for illustrative purpose
because it allows to unambiguously determine 
the point of analytical continuation we need as 
\(P^2=m_B^2+i\epsilon\).
Finally, it is 
convenient to represent the QCD calculation result 
\(\Pi_{\pi\pi}^{(O,T)}((p-q)^2,m_B^2)\) in the form of the 
dispersion relation in the variable \((p-q)^2\), 
\begin{equation}
  \Pi_{\pi\pi}^{(O,T)}((p-q)^2,m_B^2+i\epsilon)=
\frac{1}{\pi}\int\limits_{m_b^2}^\infty ds'~\frac{\textrm{Im}_{s'}
\Pi_{\pi\pi}^{(O,T)}\left(s', m_B^2 \right)}{s'-(p-q)^2-i\epsilon}\,,
 \label{eq-Psq-disp}
\end{equation}
and equate this 
to the hadronic representation in the $B$ meson channel,
(see Fig.~8)
\begin{equation}
  \Pi_{\pi\pi}^{(O,T)}((p-q)^2,m_B^2+i\epsilon)=\frac{f_B
m_B^2\langle\pi^-(p)\pi^+(-q)|O|B(p-q)\rangle_T}
  {m_B^2-(p-q)^2}+\int\limits_{s_h^B}^\infty
ds'~\frac{\rho_h^{(B)}(s')}{s'-(p-q)^2}\,,
\label{eq-Psq-disp1}
\end{equation}
where $f_B$ is the $B$-meson decay constant. 
In the ground-state contribution, we have 
\(P^2=(p-k-q)^2=m_B^2+i\epsilon\) and \((p-q)^2=m_B^2\) simultaneously,
so that the artificial momentum \(k\) disappears and we encounter
the hadronic on-shell matrix element of our interest.
After applying the duality approximation 
to the integral over excited states in Eq.~(\ref{eq-Psq-disp1})
and performing Borel transformation, we obtain the LCSR 
for the $B\to \pi\pi$ hadronic matrix element of a given
operator and topology: 
\begin{equation}
  \langle\pi^-(p)\pi^+(-q)|O|B(p-q)\rangle_T
  =\frac{1}{f_B m_B^2 \pi} \int\limits_{m_b^2}^{s_0^B}ds'~e^{(m_B^2-s')/M'^2}
\mbox{Im}_{s'}\Pi_{\pi\pi}^{(O,T)}(s'+i\epsilon,m_B^2+i\epsilon). 
\label{eq-LCSRBpipi}
\end{equation}

Note that the analytical continuations of \(\Pi_{\pi\pi}^{(O,T)}\) in the
variables \(s'\) and \(P^2\) interchange
and may be performed in inverse order as well.

\section{Annihilation with hard gluons}
\label{Sec-Hard}
\begin{figure}[t]
\begin{center}
\subfigure[]{\includegraphics[scale=0.5]{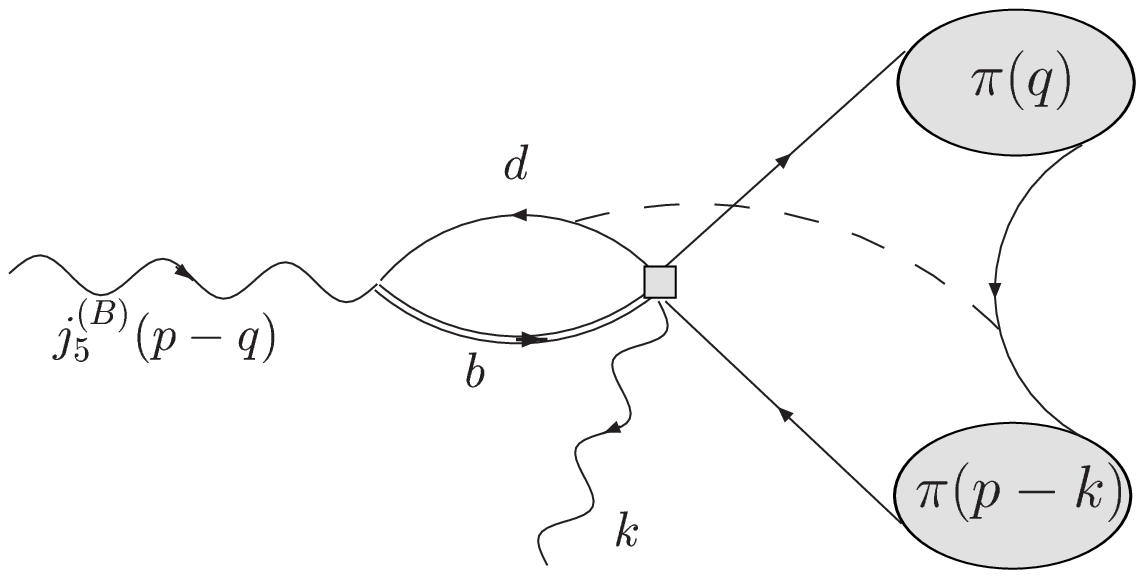}}
\hspace{0.5cm}
\subfigure[]{\includegraphics[scale=0.5]{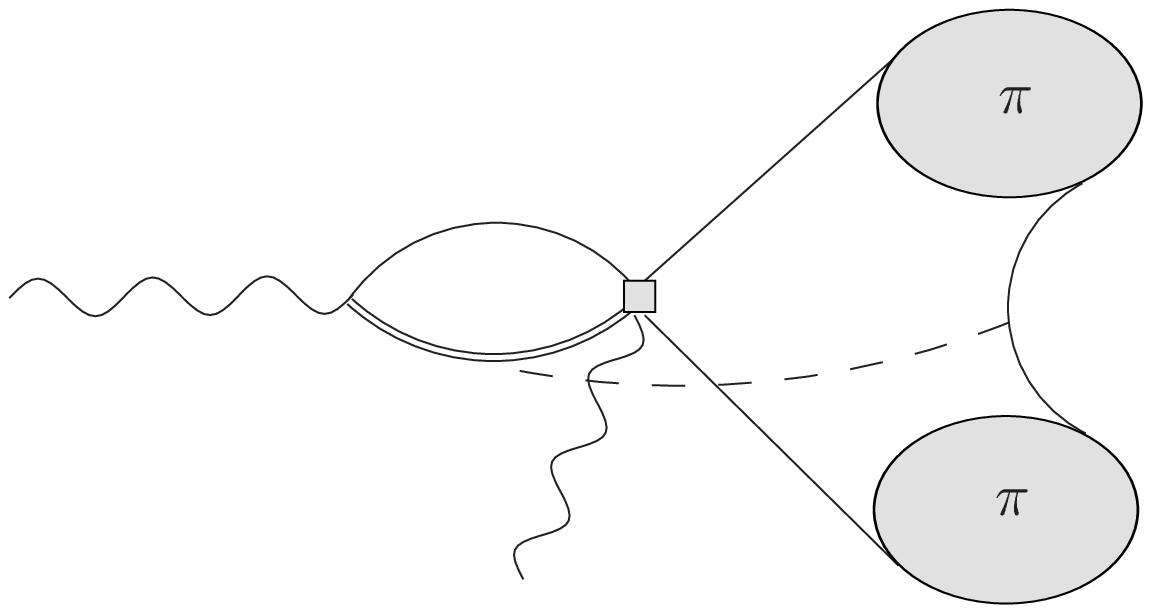}}
\end{center}
\caption{\em Diagrams  used to calculate the pion-pion
correlator in the annihilation topology.}
\label{fig-HardAnn}
\end{figure}

For the correlation function (\ref{eq-F}) with the operators
\(O_{1,2}^u\), the simplest possible diagram with 
the annihilation topology 
is the factorizable diagram in Fig.~\ref{fig-OPE_A}a.
Its contribution is expected to vanish due to the conservation
of the $V-A$ current for massless $u,d$ quarks.
In fact, a calculation of the diagram in LCSR yields a
parametrically small correction of $O(s_0^\pi/m_B^2)$ which 
is neglected within the adopted accuracy of the method
\cite{AKBpipi}, so that the result is consistent with the expectation.


We start to calculate the annihilation effect by considering the first two
\(O(\alpha_s)\) diagrams shown in Fig.~\ref{fig-OPE_A}b; they contain only the
operator \(\widetilde O_2^u\).
These two-loop diagrams depend
on four different mass/momentum scales, 
$m_b$, $(p-k)^2$ , $(p-q)^2$ and $P^2$,
which makes their direct calculation technically not feasible. 
As mentioned 
in the previous section, 
instead of calculating these diagrams directly,
we start from the pion-pion 
correlator $\Pi_{\pi\pi}^{( \widetilde{O}_2^u,A)}$  
defined in  Eq.~(\ref{eq-Pipipi-Corr}),
with the annihilation topology.  
The corresponding one-loop 
diagrams are shown in Fig. \ref{fig-HardAnn}
and we proceed with their calculation.

After contracting the quark and gluon fields, 
the long-distance part of 
$\Pi_{\pi\pi}^{( \widetilde{O}_2^u,A)}$,
at leading twist 2, reduces to
a product of two pion twist-2 DA's:
\begin{align*}
&\langle
\pi^-(p-k)|\bar{d}_a^\alpha(z)\,u_b^\beta(0)~~\bar{u}_c^\gamma(0)\,d_d^\delta(
z)|\pi^-(q)\rangle\\
&\approx~~\langle \pi^-(p-k)|\bar{d}_a^\alpha(z)u_b^\beta(0)|0\rangle\
\cdot \ 
 \langle 0|\bar{u}_c^\gamma(0)d_d^\delta(z)|\pi^-(q)\rangle\\
&=~~\Biggl(\frac{i\delta_{ab}}{12}f_{\pi}\left[(\DS{p}-\DS{k})\gamma_5\right]^
{\beta\alpha}
\int\limits_0^1 dv~e^{i v (p-k)\cdot z}\varphi_\pi(v)\Biggr)
\cdot
\Biggl(\frac{-i\delta_{cd}}{12}f_{\pi}\left[\DS{q}\gamma_5\right]^
{\delta\gamma}
\int\limits_0^1 du~e^{-i u q\cdot z}\varphi_\pi(u)\Biggr)\,,
\end{align*}
where $\alpha,\beta,\gamma,\delta$ and $a,b,c,d$ are the 
spinor and color indices, respectively.
Convoluting the above with the short-distance part,
one obtains the following expression
for the pion-pion correlator :
\begin{multline}
\label{eq-Pipipi-diagram}
\Pi_{\pi\pi}^{(\widetilde{O}_2^u,A)}\left((p-q)^2,P^2\right)
=-i\left(\frac{\alpha_sC_F}{\pi}\right)\frac{f_\pi^2m_b^2}{48}
\int\limits_0^1 du\int\limits_0^1 dv~\frac{\varphi_\pi(u)\varphi_\pi(v)}
{[(p-q)^2 u-P^2 v]^2}
\times\\
\left\{H_{\pi\pi}^d\left(u,v;(p-q)^2,P^2\right)+H_{\pi\pi}^b\left(u,v;(p-q)^2,
P^2\right)\right\}\,,
\end{multline}
where \(H_{\pi\pi}^{d}\) and \(H_{\pi\pi}^{b}\) 
represent the hard-scattering 
kernels for the diagrams  in Figs.~\ref{fig-HardAnn}a 
and \ref{fig-HardAnn}b, with the gluon attached 
to the \(d\) and \(b\) quark, 
respectively. The expressions for $H_{\pi\pi}^{d,b}$ are given 
in the appendix \ref{Sec-MessyFormula}
in terms of the standard loop integrals.
Evaluated at \((p-q)^2,P^2<0\), the  result for 
Eq.~(\ref{eq-Pipipi-diagram}) is real and finite,
in particular, the numerators in
Eq.~(\ref{eq-Pipipi-diagram}) cancel at 
the pole \((p-q)^2 u=P^2 v\).  
Furthermore, \(\Pi_{\pi\pi}^{(\widetilde{O}_2,A)}\)
contains no end-point divergences in the variables 
$u$ and $v$.
The reason why the integrals remain finite is simple:
the gluon in the diagrams in Fig.~\ref{fig-HardAnn} 
has a virtuality \(u v P^2\),
and remains perturbative unless \(u\) or \(v\) is close to \(0\), 
but this region is suppressed by the endpoint behavior 
of the pion DA's. This convergence is important for justifying 
the replacement of the initial diagrams in Fig.~\ref{fig-OPE_A}b.
Note that the latter are convergent and perturbative 
'by construction', because they contain one of-shell 
current instead  of the pion DA. 
 
For the sum rule derivation we need  the dispersion relation 
(\ref{eq-Psq-disp1}) for $\Pi_{\pi\pi}^{(\widetilde{O}_2,A)}$
in the variable \(s'=(p-q)^2\). The expression
(\ref{eq-Pipipi-diagram}) continued in this variable, 
(while keeping \(P^2\) negative), has branch cuts at $s'>0$.  
The calculation of the imaginary part is straightforward,
but involved. The result reads: 
\begin{multline}
\label{eq-Im-Pipipi}
\mathrm{Im}_{s'}
\Pi_{\pi\pi}^{(\widetilde{O}_2,A)}\left(s'+i\epsilon,P^2\right)
=i f_\pi^2 m_b^2\left(\frac{\alpha_sC_F}{24}\right)
\int\limits_0^1 du \int\limits_0^1
dv~
\frac{\varphi_\pi(u)\varphi_\pi(v)}{u v (u\,s'-v\,P^2)^3}
\\\times
\left[H_1(u,v;s',P^2)\cdot\Theta(s'-m_b^2)+H_2(u,v;s',P^2)\cdot
\Theta\left(s'-\frac{m_b^2}{\bar{u}}-v\,P^2\right)\right],\
\end{multline}
where $\bar{u}=1-u$ and 
\begin{multline*}
H_1(u,v;s',P^2)=
u v\,P^2 (m_b^2-s'+v\,P^2)^2
\log\left(\frac{v\,P^2 |m_b^2-\bar{u}s'|}{u\,s' (m_b^2-s'+v\,P^2)}\right)\\
-\left(1-\frac{m_b^2}{s'}\right)(u\,s'-v\,P^2)(m_b^2-u\,s'+v\,P^2)
     (u\,s'+v\,P^2)\\
-\left(u(m_b^2- u\,s')+v\,P^2\right)
   \left(v\,m_b^2\,P^2+u\,s'(s'-v\,P^2)\right)
\log\left|\frac{s'(u\,m_b^2+\bar{u}v\,P^2)}
               {u\,s'^2 + v\, P^2 ( m_b^2 - u\,s')}\right|\,,
\end{multline*}
\begin{multline*}
  H_2(u,v;s',P^2)=
-u v\,P^2(m_b^2-s'+v\,P^2)^2
\log\left(\frac{\bar{u}v\,P^2(m_b^2-s'+v\,P^2)}
               {u(s'-v\,P^2)|m_b^2-\bar{u}\,s'|}\right)\\
+(u\,s'-v\,P^2)\left(1-\frac{m_b^2}{\bar{u}(s'-v\,P^2)}\right)
  \left(m_b^2(u\,s'+(1-2 u)v\,P^2)-2 u(s'-v\,P^2)(u\,s'-v\,P^2)\right)\\
+\left(u(m_b^2-u\,s')+v\,P^2\right)  
  \left(v\,P^2 m_b^2+u\,s'(s'-v\,P^2)\right)
  \log\left|\frac{(s'-v\,P^2)(u\,m_b^2+\bar{u} v\, P^2)}
                 {\bar{u}(u\, s'^2 + v\,P^2( m_b^2 - u\,s'))}\right|
\,.
\end{multline*}

The expression (\ref{eq-Im-Pipipi}) at negative $P^2$ 
contains no singularities  within the integration region, 
and is therefore finite.
The two pieces proportional to \(\Theta(s'-m_b^2)\) and
\(\Theta(s'-m_b^2/\bar{u}-P^2 v)\) reflect the 
two cuts of the diagrams in Fig.~\ref{fig-HardAnn}:
the first one corresponds to 
the on-shell $b$ and $\bar{d}$ quarks emitted at the 
\(B\) current vertex; the second cut is 
less trivial and emerges when $b$- and $d$- quark
at the weak vertex are on-shell. 

Finally, according to the procedure explained 
in the previous section, we analytically continue 
$\mathrm{Im}_{s'}\Pi_{\pi\pi}^{(\widetilde{O}^u_2,A)}$ to 
the physical timelike point 
\(P^2= m_B^2+i\epsilon\), so that this function acquires 
an imaginary part \( \mathrm{Im}_{P^2}\mathrm{Im}_{s'}
\Pi_{\pi\pi}^{(\widetilde{O}^u_2,A)} \).
The imaginary part in (\ref{eq-Im-Pipipi})
naturally originates from the logarithms of $-P^2$, 
however the complexity
of this expression makes their extraction nontrivial. 
We obtain
\begin{multline}
\label{eq-Pipipi-double-im}
\mathrm{Im}_{P^2}\mathrm{Im}_{s'}\Pi_{\pi\pi}^{(\tilde{O}^u_2,A)}\left(s',m_B^
2\right)=
i\frac{\pi\alpha_s C_F}{24}f_\pi^2\, m_b^2\, m_B^2 (m_b^2-s'+v\,m_B^2)^2\\
\times\int\limits_0^1 du \int\limits_0^1 dv
\frac{\varphi_\pi(u)\varphi_\pi(v)}{(u\,s'-v\,m_B^2)^3}
\left[\Theta\left(s'-\frac{m_b^2}{\bar{u}}\right)
-\Theta\left(s'-m_b^2-v\,m_B^2\right)\right]\,.
\end{multline}
The existence of nonvanishing imaginary part in $P^2$  
is an important effect we are actually looking for, 
because in the quark-hadron
duality approximation it 
determines the strong phase of the hadronic matrix element.
Importantly, 
Eq.~(\ref{eq-Pipipi-double-im}) receives contributions   
only from the diagram
in  Fig.~\ref{fig-HardAnn}a. Physically, the effect
corresponds to the $d$ quark from the  weak decay of $b$ quark  
going on shell and annihilating with the spectator $\bar{d}$  
quark into a virtual timelike gluon. 
In the diagram in Fig.~\ref{fig-HardAnn}b 
the gluon is attached to the $b$ quark 
and such mechanism is forbidden kinematically, 
hence this diagram has no double imaginary part.
We have verified by explicit calculation
that an identical result is obtained if one does 
analytical continuation in \(P^2\) first and only then 
obtains the dispersion relation in \(s'\).

Our final result 
for the annihilation contribution with hard gluons
in twist 2 approximation is given by the sum rule 
of the type (\ref{eq-LCSRBpipi}):
\begin{equation}
\label{eq-Hard-Result}
\langle\pi^-\pi^+|\widetilde{O}_2^u|\bar{B}^0\rangle_A^{hard}
=\frac{1}{f_B m_B^2\pi}\int\limits_{m_b^2}^{s_0^B} ds'~e^{(m_B^2-s')/M'^2}
\mathrm{Im}_{s'}\Pi_{\pi\pi}^{(\widetilde{O}^u_2,A)}
(s',m_B^2+i\epsilon)\,,
\end{equation}
where the real and imaginary part are given 
by the real part of Eq.~(\ref{eq-Im-Pipipi}) 
(with the principal value of the integrals containing complex poles) 
and by Eq.~(\ref{eq-Pipipi-double-im}), respectively.

As already mentioned, the LCSR result (\ref{eq-Hard-Result})
is finite, due to 
the fact that the end-point divergence introduced by the 
gluon propagator \(1/u v P^2\) is cancelled by the pion DA's.
In the light-cone expansion of the
two-pion diagrams in Fig.~\ref{fig-HardAnn}
one formally encounters a contribution proportional to the 
two twist-3 pion DA's 
which is divergent. However, one has to keep in mind, 
that we have used these diagrams  
only as an effective replacement 
for the part of the diagrams  in the correlation function
(\ref{eq-F}).
In this function
one of the pions is interpolated by the axial-vector current,
which simply does not have a twist 3 component. 
We conclude that  the (twist 3)$\otimes$(twist 3) 
contribution to \( \Pi_{\pi\pi}^{(\widetilde{O}^u_2,A)}\) 
has no counterpart in the correlation function (\ref{eq-F})
and hence does not play any role in LCSR.

To complete the calculation of hard-gluon effects, 
we still need to consider the 
four diagrams in Fig. \ref{fig-OPE_A}c which belong 
to the $O(\alpha_s)$ part of the correlation function (\ref{eq-F}) . 
Note that similar diagrams where $B\to \pi\pi $ 
annihilation is accompanied by gluons 
exchanged within the weak vertex, also emerge  
in QCDF approach. However, they were not included in \cite{BBNS2},
where only the diagrams analogous to 
Fig. \ref{fig-OPE_A}b were taken into account.  
The reason, apart from  expected $1/m_b$ suppression, is that 
in order to describe the
two-pion state originating from a quark-antiquark pair, 
in QCDF one needs an additional  long-distance object, 
a sort of two-pion distribution amplitude, which in the local 
limit reduces to the pion form factor at timelike momentum 
transfer $m_B^2$.
In LCSR approach 
the  `formfactor-like' annihilation diagrams in Fig.~\ref{fig-OPE_A}c 
emerge as a part of OPE, hence, no new input is needed. 
However, the calculation of these diagrams 
is not possible with current methods,
because they contain two loops  and many scales. 
In fact, in this case one cannot use
as a remedy  the pion-pion correlator considered above, 
since the corresponding 
diagrams still have two loops, even if both pions  are described 
by their DA's. In order to assess this effect, remaining at the one-loop 
level, we employ a completely different 
method which is briefly described in the rest of this section.

A new type of correlation function 
is introduced with an on-shell \(B\) meson and a pion,  
while interpolating the second pion 
with a current \footnote{\, A similar but simpler version 
of this method was recently applied to $B\to \pi$ form factor in 
Ref.~\cite{KMO} (see also Ref.~\cite{FHDF}).}:
\be
\Pi^{B\pi}_\alpha(p,k,q)
= i\int d^4 y~e^{i(p-k)y}
\langle0|T\left\{j_{\alpha 5}^{(\pi)}(y),\tilde{O}_2^u(0)\right\}
|B(p-q) \pi^-(q)\rangle\,.
\label{eq-corrphiB}
\ee
We consider the part of this correlation function 
with quark contractions having annihilation topology 
and pick up only diagrams with gluon exchange 
in the weak vertex. 
They can be obtained 
from the diagrams in Fig. \ref{fig-OPE_A}c if 
the interpolating current $j_5^{(B)}$  is replaced by 
an on-shell $B$-meson. 
The correlation function (\ref{eq-corrphiB}) factorizes 
into a product of vacuum-to-pion 
matrix element (that is, a usual pion DA) 
and vacuum-to-\(B\) matrix element. The latter
is expressed via $B$ meson 
DA's \cite{GN,BenekeFeldmann}. 
As usual, the external momenta are chosen 
to provide that the virtual quarks and gluon remain 
far off-shell. For the pion DA 
we retain only twist 2 
(twist 3 vanishes in the chiral limit) whereas 
the two components $\phi^{\pm}$ of the $B$ meson DA 
are taken into account. By writing  a dispersion relation 
in the variable \((p-k)^2\) (in the pion channel), one obtains 
a sum rule for the \(B\to\pi\pi\) matrix element.
As a result, we find that all four diagrams vanish,
which means that
contributions from the ``form factor'' part of 
the annihilation mechanism starts either 
at higher twists \(\geq 4\) and/or at higher orders in 
\(\alpha_s\) and can be neglected.

\section{Annihilation with soft gluons}
\label{Sec-Soft}
The gluons exchanged between the initial \(B\)-meson and the 
final quark-antiquark state in the $B\to \pi\pi$ annihilation 
can also have small virtualities.
Diagrams with soft gluon contributions 
cannot be directly calculated  
in QCDF or PQCD, because in this case 
it is difficult to identify and separate a hard kernel. 
The soft-gluon nonfactorizable 
effects should either be neglected (arguing that they are 
$1/m_b$ suppressed) or modelled by separate nonperturbative 
parameters. 
In the LCSR approach the decay amplitude is calculated
quite differently, by matching 
the hadronic dispersion relation to the correlation function.
In the latter, the soft (low virtuality) gluons emerge in OPE
diagrams, being emitted at 
short distances and absorbed in the quark-antiquark-gluon
DA's of the pion.
For the emission topology the corresponding diagrams
are shown in Fig.~\ref{fig-OPE_E}c. 
Their effect, although formally $1/m_b$ 
suppressed turned out \cite{AKBpipi} to be of the same 
order as the $O(\alpha_s)$ effect of nonfactorizable 
hard gluons calculated from QCDF. On the other hand, 
the soft-gluon effects 
for the penguin topology (one of diagrams is shown in 
Fig.~\ref{fig-OPE_P}b) were found suppressed in LCSR \cite{KMMpeng}
with respect to the penguin diagrams with hard gluons,
indicating that the role of soft-gluon effects 
strongly depends on the topology.
Here we will calculate the 
soft-gluon diagrams in the part 
of the correlation function (\ref{eq-F})
with annihilation topology.

The two lowest-order diagrams
are shown in Figs.~\ref{fig-OPE_A}d,e 
containing an on-shell gluon  emitted from the heavy-light loop 
and absorbed in the three-particle pion DA. 
Technically these diagrams are much easier
to calculate than the annihilation diagrams with hard gluons
in Fig~\ref{fig-OPE_A}b,c.
One returns to the original method of \cite{AKBpipi}
and employs the light-cone expansion of the quark propagators in the 
external  gluon field \cite{BB}. For the diagram with a gluon emission
from the massless $d$ quark we use
\be
S_d(x,0)=-i\langle 0|T\{ d(x)\bar{d}(0) \}|0\rangle 
= \frac{\DS x}{2\pi^2 (x^2)^2}-\frac{1}{16\pi^2x^2}\int\limits_0^1 dv 
G^{\tau\rho}(vx)(\DS x \sigma_{\tau\rho}-4ivx_\tau\gamma_\rho)
+\ldots\,,
\label{eq-propD}
\ee
whereas the propagator for the massive $b$ quark is
\begin{multline}
S_b(0,x)=-i\langle 0 |T\{ b(0)\bar{b}(x) \}|0\rangle 
=\int \frac{d^4k}{(2\pi)^4}e^{ikx}\frac{\DS k +m_b}{k^2-m_b^2} 
\\
-\int \limits_0^1 dv G^{\tau\rho}(vx) \int \frac{d^4k}{(2\pi)^4}e^{ikx}
\Big[ \frac12 \frac{\DS k
+m_b}{(k^2-m_b^2)^2}+\frac{\bar{v}x_\tau\gamma_\rho}{k^2-m_b^2}  \Big]
+\ldots\,.
\label{eq-propB}
\end{multline}
In the above $\bar{v}=1-v$, $G_{\tau\rho}= 
\frac{\lambda^a}{2} G^a_{\tau\rho}$
and the fixed-point gauge for the gluon field has been adopted, 
having in mind that the hard and soft contributions 
to OPE are individually gauge-invariant.
The dots in Eqs.~(\ref{eq-propD}) and (\ref{eq-propB}) 
represent terms with derivatives and higher orders of 
the gluon field-strength tensor which we neglect. These terms  
generate pion DA's  with twist $>4$ and multiplicity $>3$, 
and their contributions to the sum rule  are suppressed 
by additional powers of the Borel parameter.

In the chiral limit for the light quarks, the contribution 
of the twist-3 quark-antiquark-gluon pion DA vanishes and 
the nonvanishing part comes from the twist 4. 
The four relevant DA's $\widetilde{\varphi}_{\parallel,\perp}$ and  
$\varphi_{\parallel,\perp}$ are defined via vacuum-pion 
matrix elements:
\begin{multline}
\langle 0| \bar u(0)i\gamma_\mu \widetilde G_{\alpha\beta}(x_3)
d(x_1) |\pi(q)\rangle = 
f_\pi \int\! {\cal D}\alpha_i~e^{-iq(x_1\alpha_1+x_3\alpha_3)}\\
\times
\Bigg[( g_{\mu\alpha}q_\beta - g_{\mu\beta}q_\alpha)
\widetilde\varphi_\perp(\alpha_i)
+ q_\mu \frac{z_\beta q_\alpha-z_\alpha q_\beta }{qz}
\Big( \widetilde\varphi_\perp(\alpha_i)+
\widetilde{\varphi}_{\parallel}(\alpha_i)\Big)\Bigg]\,,
\label{eq-tw4matr}
\end{multline}
and the one  obtained from the above with 
$i\gamma_\mu \to \gamma_\mu\gamma_5 $,
$\widetilde{G}_{\alpha\beta} \to  G_{\alpha\beta}$ and 
$\widetilde{\varphi}_{\parallel,\perp} \to \varphi_{\parallel,\perp}$.
In Eq.~(\ref{eq-tw4matr}), $\widetilde G_{\alpha\beta}= \frac 12 
\epsilon_{\alpha\beta\rho\lambda}G^{\rho\lambda}$, 
${\cal D} \alpha_i = d\alpha_1d\alpha_2d\alpha_3
\delta \left(1- \alpha_1 - \alpha_2- \alpha_3 \right)$.
The points $x_{1,2}$ are located on the light cone, 
$x_i=u_iz$,  where $u_i$ are arbitrary numbers and $z^2=0$ 
is the light-cone separation.

In our case, due to the choice of external spacelike momenta,
the regions of integration over $x,y$ are close to the light-cone, 
but $x^2,y^2,(x-y)^2$ are not exactly light-like. Therefore, 
strictly speaking, there is an ambiguity of defining 
 $z$ via $x$ and $y$ in r.h.s. of Eq.~(\ref{eq-tw4matr}). 
From the point of view of the light-cone
OPE, this ambiguity is a higher twist effect. 
Indeed, calculating the diagrams for 
different choices $z=x,y,x-y$ in Eq.~(\ref{eq-tw4matr})  
one finds that the parts proportional to $z_\alpha/(qz)$ 
in Eq.~(\ref{eq-tw4matr}) and its analog for $\gamma_\mu\gamma_5$ 
yield negligibly small contributions. Hence, only two DA's
$\varphi_{\perp}$ and $\widetilde{\varphi}_{\perp}$  
multiplying the coordinate-independent part of the matrix elements 
appear in the final answer.   

After specifying the propagators and pion DA's, 
the calculation of the diagrams in Figs.~\ref{fig-OPE_A}d,e 
is straightforward. The following expression is obtained for the 
invariant amplitude multiplying $(p-k)_\alpha$
in the correlation function :
\begin{multline}
F_{\textrm{soft}}^{(\widetilde{O}_2^u,A)}
=\frac{m_b^2 f_\pi}{16\pi^2}
\int\limits_0^1 \frac{ d\alpha_1}{-P^2\alpha_1-(p-k)^2(1-\alpha_1)} \int\limits
_0^{(1-\alpha_1)} d\alpha_3
\int\limits_0^1 dv
\int\limits_0^1 \frac{dx}{m_b^2-(p-q)^2(1-\alpha_3v)x}
\\ 
\times 
\Bigg\{\Big[
P^2(1+2x\bar{v})+3(p-q)^2(1-2\bar{v}x)\Big]\varphi_\perp(\alpha_i)-
\Big[ P^2(1-4x\bar{v})+3(p-q)^2\Big]\widetilde{\varphi}_\perp(\alpha_i)\Bigg\}\,.
\label{eq-Fasoft}
\end{multline}
Following
the procedure explained in section \ref{Sec-Method},
we match the above expression to the dispersion 
relation in the variable  \((p-k)^2\).
After applying duality and Borel transformation, we obtain
for the corresponding pion-pion correlator: 
\begin{multline}
\Pi_{\pi\pi}^{(\widetilde{O}_2^u,A),\textrm{soft}}= 
i\frac{m_b^2}{16\pi^2P^2}\int\limits_0^{s_0^\pi}ds~e^{-s/M^2}
\int\limits_0^1 d\alpha_3
\int\limits_0^1 dv \int\limits_0^1 \frac{dx}{m_b^2-(p-q)^2(1-\alpha_3v)x}
\\ 
\times 
\Bigg(\Big[
P^2(1+2x\bar{v})+3(p-q)^2(1-2\bar{v}x)\Big]\varphi_\perp(0,\bar{\alpha_3},\alpha_3
)-
\Big[
P^2(1-4x\bar{v})+3(p-q)^2\Big]\widetilde{\varphi}_\perp(0,\bar{\alpha}_3,\alpha_3)
\Bigg)
\\
\times \{1+O(s_0^\pi/P^2)\}\,,
\label{eq:PisoftA}
\end{multline}
where $\bar{\alpha}_3=1-\alpha_3$ and 
we neglected terms of \(O(s/P^2)<O(s_0^\pi/P^2)\). 

The analytical continuation
to \(P^2=m_B^2\) is then trivial and we can immediately
apply the dispersion relation in the variable $(p-q)^2$,
and subsequently, duality and the Borel transformation in 
the \(B\) channel.
Our final result for the soft-gluon annihilation contribution
to the hadronic matrix element 
in the leading twist-4 approximation reads 
\begin{multline}
\langle\pi^-\pi^+|\widetilde{O}_2^u|\bar{B}^0\rangle_A^{soft}
=i\frac{m_b^2}{16\pi^2 f_B m_B^4}\int\limits_0^{s_0^\pi}ds e^{-s/M^2}
\int\limits_{m_b^2}^{s_0^B}\frac{ds'}{s'}e^{m_B^2/M'^2-s'/M'^2}
\\
\times\int\limits_{m_b^2/s'}^1\frac{du}{u}\int\limits_{1-u}^1
\frac{d\alpha_3}{\alpha_3}
\Bigg\{\Big[m_B^2+3s'+2(m_B^2-3s')\frac{m_b^2(u-\bar{\alpha}_3)}{s'u\alpha_3}
\Big]\varphi_\perp(0,\bar{\alpha}_3,\alpha_3)
\\
-\Big[m_B^2+3s'-4m_B^2\frac{m_b^2(u-\bar{\alpha}_3)}{s'u\alpha_3}\Big]
\tilde{\varphi}_\perp(0,\bar{\alpha_3},\alpha_3)\Big]\Bigg\}
\times \{1+O(s_0^\pi/P^2)\}\,.
\label{eq-softA}
\end{multline}
In the adopted approximation, this part of the decay amplitude 
does not contribute to the strong phase. 

Finally, 
adding the hard-gluon and soft-gluon
contributions given by Eqs.~(\ref{eq-Hard-Result}) and Eq.~(\ref{eq-softA}), respectively, we complete our calculation of the parameter
$r_A^{(\pi\pi)}$ defined in Eq.~(\ref{eq-rArP}).

\section{Factorizable annihilation via $O_6$ operator}
\label{Sec-O6}
In this section we describe the calculation of the 
pion scalar form factor defined in 
Eq.~(\ref{scalarFF}). According 
to Eq.~(\ref{matrO6}), this form factor is
 needed to estimate the factorizable part of 
the $B\to \pi\pi$ matrix element of the operator $O_6^d$ 
with annihilation topology.

In QCDF only part of this 
contribution was taken into account, 
namely, the annihilation diagrams 
with the gluon exchange in the final
state (see Figs. 4a,b in Ref.~\cite{BBNS2}). Clearly, this is not 
a complete answer, because the annihilation 
into two pions via the scalar current starts at 
zeroth order in $\alpha_s$. In addition, there are 
$O(\alpha_s)$ diagrams 
where a hard gluon is exchanged in the final state 
at the vertex  of the scalar current.
All these contributions lie beyond 
the usual QCDF approximation: being formally $1/m_b$ 
suppressed, they also do not allow a factorization with 
a hard kernel. One has to parameterize them
with a separate nonperturbative parameter. 

We use a different approach, 
considering the scalar  pion form factor  
$\langle \pi^+(p)\mid \bar{q}q \mid \pi^-(q)  \rangle$
($q=u,d$) as a separate object and obtaining it 
from LCSR, following the same method as in \cite{ffpion}.
The calculation is done 
at spacelike momentum transfer $P^2<0$, 
analytically continuing the result 
to large timelike $P^2=m_B^2$. 

\begin{figure}[t]
\begin{center}
\subfigure[]{\includegraphics[scale=0.5]{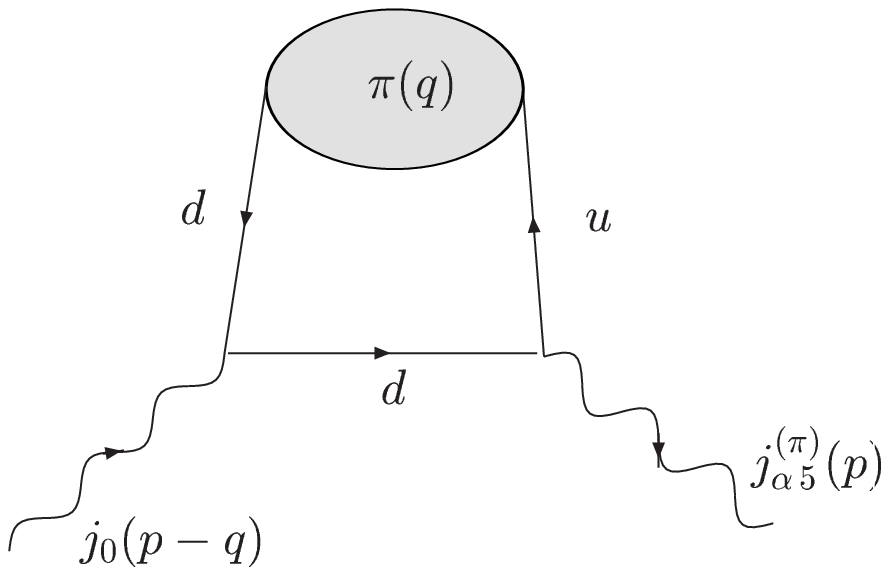}}
\hspace{0.5cm}
\subfigure[]{\includegraphics[scale=0.5]{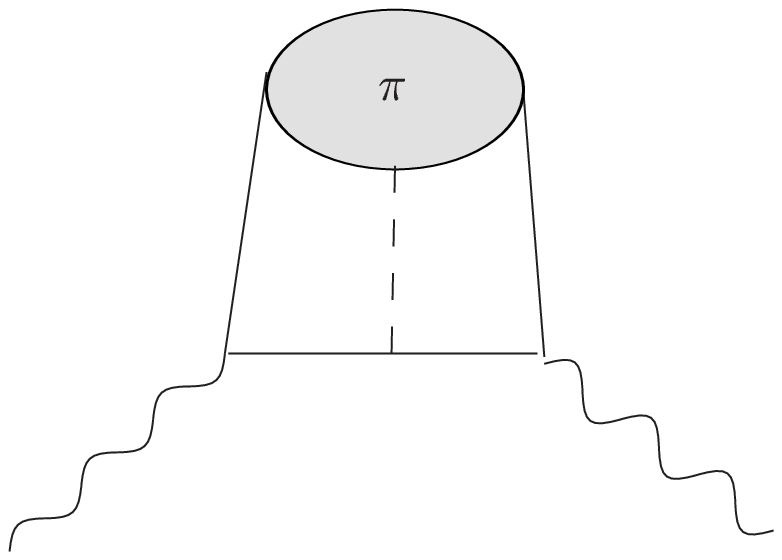}}\\
\subfigure[]{\includegraphics[scale=0.5]{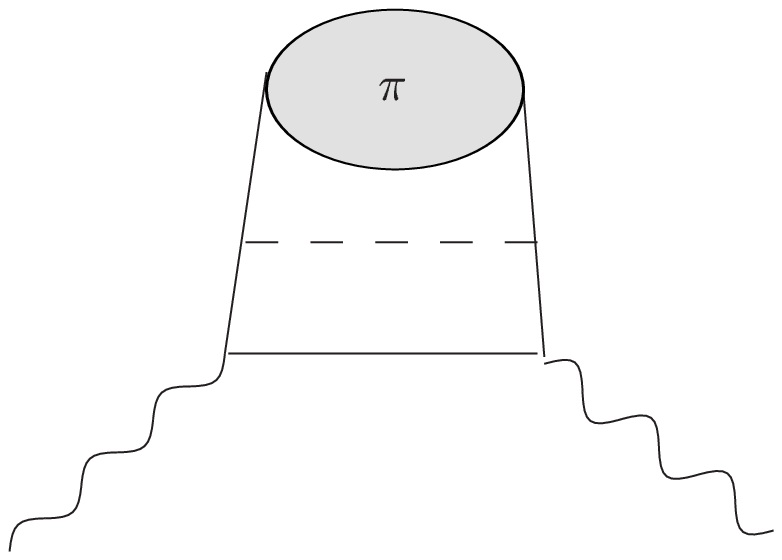}}
\hspace{0.5cm}
\subfigure[]{\includegraphics[scale=0.5]{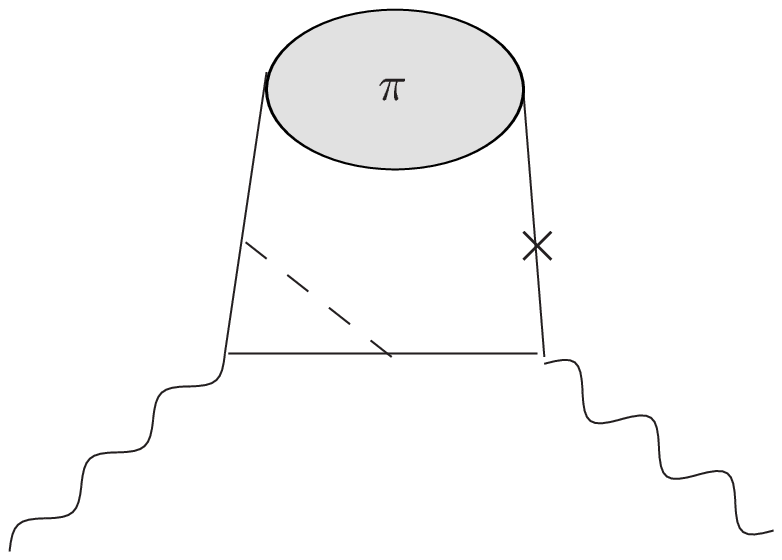}}\\
\end{center}
\caption{ Diagrams of the correlation function
used to derive LCSR for the pion
scalar form factor; \(j_0=\bar{d}d\) is the scalar quark current.}
\label{fig-scalFF}
\end{figure}
One introduces the correlation function
\be
T_\alpha(p,q) = i\int d^4 x e^{ipx}\langle 0 |
T\{j_{\alpha 5}^{(\pi)}(x) \bar{d}d(0)\}|\pi^-(q)\rangle\,,
\label{corr_scalFF}
\ee
where $j_{\alpha 5}^{(\pi)}$ is the same 
axial-vector current as in Eq.~(\ref{eq-F}). 
Inserting the complete set of states with the pion
and axial meson quantum numbers
between the currents in Eq.~(\ref{corr_scalFF}),
one obtains for the ground-state pion
contribution: 
\be
T^{(\pi)}_\alpha(p,q)=\frac{if_\pi F_\pi^S(P^2)}{m_\pi^2-p^2}p_\alpha\,.
\label{eq:scalar_pion}
\ee
At large spacelike $p^2$ and $P^2=(p-q)^2$ 
the correlation function can be expanded
near the light-cone and expressed 
via pion DA's. The OPE starts from the diagram 
in Fig.~\ref{fig-scalFF}a which corresponds to 
the ``soft'' or end-point mechanism 
of the pion-to-pion transition. 
In addition, there is a diagram  with quark-antiquark gluon DA's 
 (Fig.~\ref{fig-scalFF}b) and $O(\alpha_s)$ diagrams, some of them
shown in Figs.~\ref{fig-scalFF}c,d . 
   
To calculate the leading-order diagram 
in Fig.~\ref{fig-scalFF}a,
the free $d$-quark propagator is inserted and pion 
quark-antiquark DA's are factorized out. 
Only the structure proportional 
to $p_\alpha$ is relevant. In the chiral limit 
(at $q^2=m_\pi^2=0$) the twist 2 contribution vanishes, hence
the expansion starts from the twist 3. The answer reads:
\be
T_\alpha(p,q)=i\mu_\pi f_\pi\int\limits_0^{1} 
\frac{du}{-p^2u-P^2\bar{u}} \left(\varphi_P(u)-
\frac16\frac{d}{du}\varphi_\sigma(u)\right)p_\alpha\,,
\label{eq:LOscalar_pion}
\ee
where $\varphi_{P,\sigma}$ are the standard twist-3 
DA's of the pion, normalized to 
$\mu_\pi=m_\pi^2/(m_u+m_d)$. 
In addition, we find that the diagram in Fig.~\ref{fig-scalFF}b, 
with a low-virtuality (soft) gluon 
absorbed in the quark-antiquark-gluon DA , 
does not contribute to the relevant
Lorentz structure $\sim p_\alpha$ in the chiral limit. 

Furthermore, the experience with the LCSR for the 
pion vector (e.m.) form factor 
\cite{ffpion} tells that even at rather large momentum transfers 
of a few GeV$^2$ the $O(\alpha_s)$ diagrams in 
Fig.~\ref{fig-scalFF}c,d  are subdominant. Therefore, in what 
follows we simplify our calculation neglecting their contributions.

Equating the result (\ref{eq:LOscalar_pion}) for the
correlation function to the hadronic 
dispersion relation in the variable \(p^2\), we retain only the pion
contribution  (\ref{eq:scalar_pion})
and replace the sum over heavier states 
by a quark-hadron duality estimate.
For that one has to write the expression 
(\ref{eq:LOscalar_pion})
as a dispersion relation in the variable $s=p^2>0$
by substituting the integration variable: $u=-P^2/(s-P^2)$. 
After Borel transformation we obtain the LCSR 
for the pion scalar form factor valid at large spacelike $P^2<0$. 
Continuing it to large timelike $P^2=m_B^2$, we get
\be
F_\pi^S(m_B^2)=-\frac{\mu_\pi}{m_B^2}M^2
\left(1-e^{-s_0^\pi/M^2}\right)\left(\varphi_P(u)-
\frac16 \frac{d}{du}\varphi_\sigma(u)\right)\Bigg|_{u=1}\Bigg
(1+O(s_0^\pi/m_B^2)\Bigg)\,
\label{eq:scalar_pionFFSR}
\ee
with an evident end-point dominance.
Note that $F_\pi^S(m_B^2)$ has no imaginary part 
in the adopted approximation. It will appear at $O(\alpha_s)$, 
hence, we expect the strong phase in the form factor 
to be subdominant. 

Substituting Eq.~(\ref{eq:scalar_pionFFSR})
in Eq.~(\ref{matrO6}), we obtain the 
estimate for the parameter $R_A^{(\pi\pi,6)}$.
Containing a chiral enhanced factor, 
and having no $\alpha_s$-suppression, this effect 
is expected to be important. On the other hand,
the annihilation effects via \(O_{5,6}\) operators 
multiply small Wilson coefficients
\(c_5\) and \(c_6\) in the amplitude 
(\ref{eq-3456Bpipi}), therefore in what follows we only retain 
$R_A^{(\pi\pi,6)}$ and neglect nonfactorizable 
corrections parameterized by \(r_A^{(\pi\pi,6)}\) and
\(r_A^{(\pi\pi,5)}\).

\section{Numerical estimates}
\label{Sec-Numerics}
\newcommand{\GeV}{~\textrm{GeV}}
\newcommand{\MeV}{~\textrm{MeV}}
For the numerical evaluation we adopt the same input 
in all sum rules derived and used in this paper, 
including LCSR (\ref{eq-Hard-Result}), (\ref{eq-softA})
and (\ref{eq:scalar_pionFFSR}) for the matrix elements in the 
annihilation topology,  and the LCSR for the \(B\to \pi\) form factor 
\(f^+_{B\pi}\) \cite{Bpi} determining the 
factorizable amplitude \({\cal A_{\pi\pi}}\). 
Finally, \(f_B\) is substituted by the 
corresponding two-point sum rule, 
so that the uncertainties of the input parameters 
partly cancel in the ratios.

In the LCSR for $B\to \pi$ 
form factor \cite{Bpi} the one-loop pole mass $m_b$ 
of the $b$ quark is used.
For consistency, we adopt the same mass 
for the other sum rules.  The current interval 
$\bar{m}_b(\bar{m}_b)= 4.25 \pm 0.15$ GeV \cite{PDG}
converted into the one-loop pole mass yields  
$m_b=4.7\pm 0.1$ GeV . Furthermore,
we take $\alpha_s(m_Z)=0.1187$ \cite{PDG}, 
evolved to lower scales
at two-loop order, so that e.g., $\alpha_s(m_b/2)=0.284$.
For the factorization and renormalization scale of 
the scale-dependent input parameters, 
we use \(\mu_b=\sqrt{m_B^2-m_b^2}\simeq 2.4\) GeV,
numerically close to $m_b/2$.
Being of order of the Borel parameter $M'$ in the B-meson 
channel, this scale reflects the average virtuality 
in the correlation function. For consistency, we normalize also the Wilson
coefficients in \(H_{\textrm{eff}}\) at the same scale \(\mu_b\). For the
latter, we use the NLO results obtained in the NDR scheme 
adopted in our calculation:\\[0.5cm]

\begin{centering}
\begin{tabular}{c|ccccccc}
\(\mu\) & \(c_1\) & \(c_2\) & \(c_3\) & \(c_4\) & \(c_5\) & \(c_6\)
& \(c_{8g}\) \\
\hline
2.4 GeV & 1.124& -0.272& 0.020& -0.037& 0.010& -0.060& -0.166\\
4.8 GeV & 1.073& -0.174& 0.013& -0.034& 0.009& -0.038& -0.149\\
\end{tabular}\\[0.5cm]
\end{centering}
In our uncertainty estimates, we include the 
variation of the unified scale up to $2\mu_b$. 

Furthermore, for the condensate densities used in the sum rules, we 
take
\(\langle\bar{q}q\rangle(1\GeV)=(-0.240\pm0.010\GeV)^3\), 
so that \(\mu_\pi(1\GeV)=1.61\pm 0.20\GeV\);
\(\langle\frac{\alpha_s}{\pi} G^2\rangle=0.012\pm0.006~\GeV^4\) and
\(\langle\bar{q}Gq\rangle=(0.8\pm 0.2\GeV)\times\langle\bar{q}q\rangle(1\GeV)\), the
scale dependence of the quark-gluon condensate density
being negligible. The remaining parameters in the 
\(B\) channel are: the interval of the Borel parameter \(M'^2=10\pm 2~\GeV^2\) 
and the duality threshold \(s_0^B=35 \mp 2\GeV^2\)
chosen as in \cite{AKBpipi}.
Concerning the parameters related to the pion, 
we use the experimental value
\(f_\pi=131\MeV\) \cite{PDG}, and take the
Borel interval \(M^2=1.0^{+0.5}_{-0.2}\GeV^2\) as well as the duality
threshold \(s_0^\pi=0.7\GeV^2\) determined from the two-point 
sum rule for \(f_\pi\) \cite{SVZ}.

Finally, the pion light-cone DA's deserve a separate
discussion. Their definitions, asymptotic 
forms and nonasymptotic parts used in our calculation
can be found, e.g., in the appendix B of ref.~\cite{BK}.

In the twist-2 pion DA $\varphi_\pi(u)$ , 
we include nonasymptotic effects 
encoded by the Gegenbauer moments \(a_2^\pi\) and \(a_4^\pi\).
These parameters have recently been estimated in Ref. \cite{BZfit}
by fitting the LCSR result for $B\to \pi$ form factor to the 
data on $B\to \pi l \nu$ decay distribution:
\be
a_2^\pi(1\GeV)=0.1\pm 0.1, ~~a_4^\pi(1\GeV)\geq -0.07\,.
\ee  
We combine this range with the  
constraint \cite{Bakulev} obtained from the analysis 
of the $\pi \gamma^* \gamma$ form factor 
\be
a_2^\pi(1\GeV)+a_4^\pi(1\GeV)=0.1\pm 0.1\,.
\ee  
The resulting intervals 
\be
a_2^\pi(1\GeV)=0 \div 0.27\,,~~   
a_4^\pi(1\GeV)=-0.07 \div 0.20\,,
\label{eq-interv}
\ee
are consistent with other estimates
in the literature. Note that Eq.~(\ref{eq-interv})
does not exclude 
a purely asymptotic DA. 
Both Gegenbauer moments are independently varied
around middle values of the intervals (\ref{eq-interv})
to estimate the corresponding uncertainty.
Finally, for the twist-3 and 4 pion DA's, 
the normalization constants
and nonasymptotic parameters are taken as:  
\(f_{3\pi}\mbox{(1 GeV)}=0.0035\pm 0.0018\GeV^2\),
\(\omega_{3\pi}\mbox{(1 GeV)}=-2.88\)  
and \(\delta_\pi\mbox{(1 GeV)}=0.17 \pm0.05~\GeV^2\),
\(\epsilon_\pi\mbox{(1 GeV)}=0.5\), respectively.

Using the input specified above, we reevaluate the result of \cite{KRWWY} for
the
\(B\to\pi\) form factor:
\begin{equation}
f^+_{B\pi}(0)=0.26\pm 0.02_{[a_{2,4}^\pi]}\pm 0.03_{[param]}\,,
\label{eq-fbpinum}
\end{equation}
where the uncertainties 
induced by Gegenbauer moments and 
by other sum rule parameters (the latter 
are added up in quadrature) are shown separately.
This estimate agrees well with the most 
recent calculation of this form factor in ref.~\cite{BZ} 
where also a small twist-3 NLO correction is taken into account. 
In section 2 we added both uncertainties linearly
to be on the conservative side. In what follows, all errors are added
quadratically.

Using the sum rules for hard and soft gluon contributions
presented in sections \ref{Sec-Hard} and \ref{Sec-Soft},
we first estimate the ratio $r_A^{(\pi\pi)}$.
We find that both hard-gluon and soft-gluon 
annihilation contributions, being comparable in magnitude,   
are numerically very small 
and partly cancel each other, so that 
\be
r_A^{(\pi\pi)}= \left[-0.67^{+0.47}_{-0.87}+
i\left(3.6^{+0.5}_{-1.1}\right)\right]\times 10^{-3}\,.
\label{eq-rapipi_res}
\ee

On the other hand, the factorizable annihilation
via quark-penguin operator $O_6$ produces
a considerably larger hadronic matrix element:
\be
R_A^{(\pi\pi, 6)}=0.23^{+0.05}_{-0.08}\,.
\ee
However, the small Wilson coefficient
reduces the effect in  the decay amplitudes
to the same level as for the annihilation 
via the current-current operator: $(c_6+c_5/3)R_A^{(\pi\pi, 6)}\sim c_1
r_A^{(\pi\pi)}$.
As mentioned at the end of sect. \ref{Sec-O6}, we therefore neglect both
factorizable and nonfactorizable \(O(\alpha_s)\) corrections to 
the annihilation via the operators $O_5$ and $O_6$.
In general, the contributions of annihilation amplitudes
are found at the same level as the other nonfactorizable 
effects estimated from LCSR in 
Refs.~\cite{AKBpipi,KMU,KMMpeng}. 

Having at hand  the new estimates of $r_A^{(\pi\pi)}$ and 
$R_A^{(\pi\pi, 6)}$, we now update the phenomenological 
analysis of $B\to \pi\pi$ channels,  with all nonfactorizable 
parts of the amplitudes calculated from LCSR, except the  
emission with hard gluons which is estimated using  
QCDF. To this end, we recalculated the previous 
LCSR predictions using the current input, 
which has only slightly changed.
For the nonfactorizable emission, we obtain
\begin{equation}
r_E^{(\pi\pi)}=\Big[\left(1.8^{+0.5}_{-0.7}\right)\times
10^{-2}\Big]_\textrm{soft}
+\Big[\left(1.3^{+5.6}_{-5.2}\right)\times 10^{-2}+
i\left(-4.7^{+1.1}_{-0.3}\right)\times 10^{-2}\Big]_\textrm{hard}\,,
\end{equation}
where the soft-gluon part is obtained from \cite{AKBpipi} and
the hard-gluon contribution is estimated
using QCDF \cite{BBNS1,BBNS2} with the default value and error for the
parameter describing the twist-3 hard-spectator diagrams. 
For the quark-penguin
operators with scalar-pseudoscalar Dirac structure, 
we have found that the
soft-gluon emission contribution
vanishes in twist \(\leq 4\), and for the hard-gluon part
we again use QCDF:
\begin{equation}
r_E^{(\pi\pi,6)}=[\left(-2.7\pm 0.4\right)\times 10^{-2}]_\textrm{hard}\,.
\end{equation}

The penguin-topology effects are 
calculated from LCSR obtained in Refs.~\cite{KMU,KMMpeng}.
The resulting ratios to the factorizable amplitude are: 
\begin{alignat}{2}
r_{P_q}^{(\pi\pi)}&=\left[0.11^{+0.02}_{-0.36}
+i\left(1.1^{+0.2}_{-0.1}\right)\right]\times 10^{-2}\,,&
~~
r_{P_c}^{(\pi\pi)}&=\left[-0.18^{+0.06}_{-0.68}
+i\left(-0.80^{+0.17}_{-0.08}\right)\right]\times 10^{-2}\,,
\nonumber\\
r_{P_b}^{(\pi\pi)}&=\left(0.93^{+0.09}_{-0.65} \right)\times 10^{-2}\,,&
~~r_{8g}^{(\pi\pi)}&=-\left(3.8^{+1.3}_{-0.4}\right)\times 10^{-2}\,,
\label{eq-numpeng}
\end{alignat}
and the modified penguin parameters are 
$\bar{r}_{P_{q,c,b}}^{(\pi\pi)}\simeq  
r_{P_{q,c,b}}^{(\pi\pi)}-(\alpha_s C_F)/(36\pi)$;
for brevity we do not show the corresponding numbers.

Having specified all parameters entering 
the decomposition of $B\to\pi\pi$ amplitudes 
in Eqs.~(\ref{eq-decomptot})
and (\ref{eq-decomp1}), we calculate
the branching ratios and direct CP-asymmetries, using
the values of $B$ meson lifetimes from \cite{HFAG},
and the relevant CKM parameters from \cite{CKMfitter}.
In particular, we adopt  
\(|V_{ub}|=(4.22\pm 0.26)\cdot 10^{-3}\) 
(the errors added in quadrature) and use  
a representative interval \(\gamma=(58.6\pm 10)^\circ\).
The results are:
\begin{eqnarray}
BR(B^+ \to \pi^+ \pi^0)&=&\left(6.7^{+1.8+0.9}_{-1.5-0.8}
\right)\times \!10^{-6} \nonumber \\ 
BR(B^0 \to \pi^+ \pi^-)&=& 
\left(9.7^{+2.3+1.2}_{-1.9-1.2}
\right)\times \!10^{-6} \nonumber \\
BR(B^0 \to \pi^0 \pi^0)&=& 
\left(0.29^{+0.24+0.07}_{-0.12-0.07}
\right)\times \!10^{-6}\,,
\label{eq-BpipiBRs}
\end{eqnarray}
where the errors represent the variation of the LCSR parameters and of the
CKM factors, respectively.
The direct \(CP\) asymmetries are
presented in Table \ref{tab-CP}.
\begin{table}
\begin{centering}
\begin{tabular}{l|ccc}
 & \(a_{CP}^{dir}(B^+ \to \pi^+ \pi^0)\) & \(a_{CP}^{dir}(B^0\to\pi^+\pi^-)\)
 & \(a_{CP}^{dir}(B^0\to\pi^0\pi^0)\) \\
\hline
BaBar & \(-0.01\pm0.10\) & \(0.09\pm 0.16\) & \(0.12\pm 0.56\) \\
Belle & \(0.02\pm0.08\) & \(0.56\pm 0.14\) & \(0.44\pm 0.56\) \\
Average& \(0.01\pm0.06\) & \(0.37\pm0.10\) & \(0.28\pm0.40\)\\
\hline
This work & 0 & \( -0.04\pm 0.01\pm 0.01\) & 
\(0.70_{-0.29-0.08}^{+0.19+0.08}   \) \\
\end{tabular}\\[0.5cm]
\end{centering}
\caption{Direct \(CP\)-asymmetries from experiment 
(all numbers taken from \cite{HFAG}; the errors 
added in quadrature) compared with the  
LCSR predictions. \label{tab-CP}}
\end{table}
For completeness, we also calculate the amplitudes
\(T,P\) and \(C\) defined in (\ref{eq-PCT2})
parameterizing them as in \cite{Buras_etal}:
\be
x\,e^{i\Delta}=\frac{C_{\pi\pi}}{T_{\pi\pi}}\,,\qquad
d\,e^{i\Theta}=-\frac{P_{\pi\pi}}{T_{\pi\pi}}\,.\\
\ee
We obtain:
\begin{alignat}{2}
x&=0.29^{+0.15}_{-0.09}\,,          
&\quad \Delta&=(-21^{+9}_{-7})^\circ\,,\nonumber\\
d&=0.22^{+0.02+0.01}_{-0.03-0.01}\,,
&\quad \Theta&=(-173 \pm 1)^\circ\,,
\label{eq-amplnum}
\end{alignat}
where the second error in \(d\) stems from the uncertainty of
\(|V_{ub}|\).

We find that the general picture does
not qualitatively deviate from the naive 
factorization considered in sect~2. Although the nonfactorizable 
contributions substantially enhance the small  
$ B^0\to\pi^0\pi^0$ amplitude predicted in naive factorization,
the disagreement between theory and experiment for 
the branching ratio of this decay remains. Our
prediction (\ref{eq-amplnum}) for the amplitudes is inconsistent
with the fit \cite{Buras_etal} to  data. As already discussed in sect.~2, one possible 
interpretation of this disagreement is
a missing contribution to the \(I=0\) parts of the amplitudes
\(T_{\pi\pi}\), \(C_{\pi\pi}\) and \(P_{\pi\pi}\).

\section{Comparison with QCD factorization}
\label{Sec-Comparison}

Let us first investigate the behavior of the annihilation
amplitudes obtained from LCSR  in the heavy quark limit,
by making standard substitutions in the sum rules:
\begin{equation}
  m_B=m_b+\bar\Lambda,\quad s_0^B=m_b^2+2 m_b\omega_0,\quad M'^2=2
m_b\tau,\quad f_B=\hat{f}_B/\sqrt{m_b}\,,
\end{equation}
where \(\bar\Lambda,\omega_0,\tau\) and \(\hat{f}_B\) are \(m_b\)-independent
parameters \footnote{\,Note that the scale 
$\tau$ has to be large in comparison with 
$\Lambda_{QCD}$, allowing one to use the power expansion  
in $\Lambda_{QCD}/\tau$, so that QCD sum rules  remain valid
in the heavy quark limit 
(the well known example is the sum rule for $f_B$ in 
HQET \cite{HQETSR}).}. 
At $m_b\to \infty$, the factorizable amplitude 
${\cal A}_{\pi\pi}$ defined in Eq.~(\ref{Afact}) 
scales as $m_b^{1/2}$.
Expanding  in \(1/m_b\)   
the annihilation contribution with hard gluons given 
by Eq.~(\ref{eq-Hard-Result}) 
and dividing it by 
${\cal A_{\pi\pi}}$, 
we reduce the real part of the 
diagram in Fig.~\ref{fig-HardAnn}a 
(gluon emitted by the light quark) to
\begin{equation}
  r_{A, \textrm{hard}}^{(\pi\pi)}\sim\frac{\Lambda}{m_b}
\int\limits_0^1 du \frac{\varphi_\pi(u)}{u^2}\int\limits_0^1 
\,dv \frac{\varphi_\pi(v)}{v} +...\,,
\label{eq-rAlimit}
\end{equation}
where $\Lambda$ is a generic energy scale not related to $m_b$
and higher powers in $1/m_b$ are denoted by ellipses.
Although formally suppressed by \(1/m_b\), this 
expression recovers the logarithmic 
divergence at \(u\to 0\), of the 
annihilation contribution in QCDF. Importantly, in LCSR this divergence 
occurs only
at $m_b\to \infty$. Moreover, as the numerical analysis in sect.~\ref{Sec-Numerics}
shows, at finite $m_b\sim 5$ GeV the logarithms
originating from the end-point region  
(at $m_b\to\infty$  they reduce to $\mbox{Log}(m_b/\tau)$ and 
 $\mbox{Log}(m_b/\omega)$)  
do not produce an enhancement of the  annihilation contribution.
Furthermore, analyzing LCSR in the heavy quark limit, 
we observe that 
the phase generated by the diagram in 
Fig.\ref{fig-HardAnn}a,  as well as the    
contribution of the gluon emission from the heavy quark
(Fig.\ref{fig-HardAnn}b), 
remain finite at $m_b\to \infty$, being of the same \(O(1/m_b)\). 
Other contributions that we have calculated 
from LCSR, the soft-gluon part of $r_A^{(\pi\pi)}$  
and the factorizable annihilation with $O_6$ given 
by $R_A^{(\pi\pi,6)}$, are also finite, being
suppressed by an additional power of 
$1/m_b$ with respect to Eq.~(\ref{eq-rAlimit}).

The origin of the end-point
divergence in QCDF and the reason why it is absent in the LCSR approach
can be readily understood. The power counting in QCDF implies that 
the momentum of the light quark in the $B$ meson has to vanish, i.e. 
the propagator of the light quark shrinks to a point \cite{BBNS1,BBNS2}. 
This is different in LCSR, since  the $B$ meson
is effectively replaced by the spectral density of the heavy-light quark loop
integrated over the duality interval $m_b^2 <s'< s_0^B$, 
and thus the momentum of the light quark is non-vanishing. In other 
words, the endpoint singularity is regularized by the typical momentum of 
the light quark. 

To demonstrate this, let us slightly modify  QCDF
by convoluting the annihilation hard kernels 
with the standard $B$-meson DA. We consider 
the annihilation diagrams in Fig.\ref{fig-HardAnn}, 
where instead of the current 
$j_5^{(B)}$, an on-shell $B$ meson is inserted,
represented by  its DA. The result for the annihilation amplitude 
$A_1^i$ (see the definition in \cite{BBNS2})
in the twist 2 approximation 
reads:
\begin{multline}
A_1^i=\pi\alpha_s \int\limits_0^\infty d\omega\phi^+_B(\omega)
\int\limits_0^1du\,\varphi_\pi(u)\int\limits_0^1dv\,\varphi_\pi(v)
\\
\times
\Bigg[ \frac{1}{\bar{u}v\Big( \bar{u}-\omega/m_B\Big)} +
\frac{\bar{u}+\omega/m_B}{\bar{u}v\Big(1-(u-\omega/m_B)
(\bar{v}-\omega/m_B)\Big)}
\Bigg]\,,  
\label{eq-modelAi}
\end{multline}
where the $B$-meson DA $\phi^+_B(\omega)$ is normalized as: 
$\int_0^\infty d\omega\phi^+_B(\omega)=1$. 
Neglecting the spectator quark momentum $\omega$, that is, replacing 
$\phi^+_B(\omega) \to \delta(\omega)$,
one recovers the expression for  $A_1^i$  
given in \cite{BBNS1,BBNS2} (see also \cite{annCZ}),
with an end-point divergence in the first term in brackets
(corresponding to the diagram with gluon emission 
from the light quark).
However, it is generally expected (see e.g., \cite{GN}) that 
$\phi^+_B(\omega)\to 0$ at $\omega\to 0$. The integral 
in Eq.~(\ref{eq-modelAi}) then converges (taken as a principal
value) yielding $\mbox{Log}(m_B/\lambda_B)$, 
$\lambda_B$ being the size of the 
region in $\omega$ where the function $\phi^+_B(\omega)$ 
dominates. Simultaneously, this expression acquires an 
imaginary part, 
\footnote{\,The appearance of an imaginary part due to the momentum 
of spectator quark was noticed already in \cite{PQCD} 
while discussing the differences between QCDF and 
PQCD.}  
due to the pole in the integration region 
at $\bar{u}=\omega/m_B$.

Employing a realistic model of $\phi^+_B(\omega)$,
e.g., the one suggested in \cite{GN}:
\be
\phi^+_B(\omega)=\frac{\omega}{\lambda_B^2}e^{-\omega/\lambda_B}\,, 
\label{eq-modelphiB}
\ee
it is easy to calculate Eq.~(\ref{eq-modelAi})
 numerically and to estimate
the corresponding parameter $r_A^{(\pi\pi)}$ (see Appendix A for
the relation between this parameter and $A_1^i$).
The result turns out small, with both real and imaginary parts 
at the level of $\sim 1\%$, that is, roughly of the same size 
as the LCSR estimate obtained in the previous section.
The model of annihilation represented by Eq.~(\ref{eq-modelAi}) 
is rather crude, 
because the transverse momenta of the quarks in $B$ meson are 
neglected, but their account could  not   
qualitatively change the result. Thus, from the point of view of 
the LCSR approach,  the end-point divergence 
in the annihilation diagrams in QCDF  originates
in the hard-scattering approximation and the $m_b\to \infty$ limit.

In the phenomenological analysis of $B\to\pi\pi$ done in 
Ref. \cite{BBNS2},
a model for the annihilation diagrams  was used, 
replacing all divergent integrals by a generic
logarithm:
\be
\int\limits_0^1\frac{dy}{y}\to X_A=
\left (1+\rho_Ae^{i\varphi_A }\right)\ln\frac{m_b}{\Lambda_h}\,,
\label{fitQCDF}
\ee
where $\Lambda_h=0.5$ GeV, $\rho_A<1$ and the phase
$\varphi_A$ is arbitrary. With this model  
the effective annihilation 
coefficients $({\cal B}_{\pi\pi}/{\cal A}_{\pi\pi}) b_i$,
(where ${\cal B}_{\pi\pi}=i(G_F/\sqrt{2}) f_B f_\pi^2$) 
entering the decay amplitudes have been  
estimated \cite{BBNS2}. Using our results and the relations
given in Appendix \ref{Sec-ri_vs_ai}, we can also obtain these coefficients: 
\begin{alignat}{2}
\frac{{\cal B}_{\pi\pi}}{{\cal A}_{\pi\pi}}b_1
&=\left[-0.15^{+0.11}_{-0.19} +i\left(0.82^{+0.11}_{-0.27}\right)\right]\times 10^{-2}\,,&~~
b_2&=\frac{c_2}{c_1}b_1\,,\nonumber\\
\frac{{\cal B}_{\pi\pi}}{{\cal A}_{\pi\pi}}b_3
          &=\left[-1.3^{+0.7}_{-0.3} + i\left(0.015^{+0.002}_{-0.008}\right)\right]
            \times10^{-2}\,,&~~
b_4&\simeq\frac{c_4}{c_1} b_1\,.
\label{eq-bi}
\end{alignat}
The above estimates, especially for $b_1$,  differ from  
the ones presented  in \cite{BBNS2} for the default value 
$\rho_A=0$ in Eq.~(\ref{fitQCDF}). In fact, a numerical agreement 
is not anticipated, because the two sets of $b_i$'s  
originate from two different methods. Moreover, 
one cannot expect that LCSR predictions for $b_i$ allow 
a parameterization with a single complex parameter $X_A$. 
Important is that both the default values 
of the annihilation coefficients  
in \cite{BBNS2} and our predictions in Eq.~(\ref{eq-bi}) 
are very small in comparison with the 
factorizable amplitude. Hence, LCSR is in a qualitative agreement 
with QCDF, if the annihilation effects in the latter are represented
by moderate logarithms.

To complete the comparison with other methods, let us 
briefly discuss PQCD  \cite{PQCD}. In this approach,
all $B\to\pi\pi$  amplitudes as well as the 
$B\to \pi$ form factor are represented by the diagrams 
with $O(\alpha_s)$ hard-scattering kernels 
and meson wave functions. Another 
distinctive feature of PQCD concerns 
nonvanishing transverse momenta of partons in mesons, 
which can only be introduced in a model-dependent way.
Hence, the annihilation amplitudes in this approach 
are protected from 
the end-point divergences and acquire imaginary parts.
There is however a basic difference between LCSR and PQCD
approaches to $B\to\pi\pi$. 
In LCSR the diagrams at $O(\alpha_s)$
are subleading and numerically suppressed, 
justifying the perturbative expansion within OPE. Moreover, the 
higher-twist soft-gluon diagrams of $O(1/m_b)$ are as important 
as the $O(\alpha_s)$ effects.
Importantly, the dominant part of the $B\to \pi$ form factor
in LCSR is ``soft'' and has no relation to $\alpha_s$. 
In PQCD, the whole form factor (hence, the factorizable
part of the $B\to\pi\pi$ amplitude) is assumed to be of $O(\alpha_s)$. 
Furthermore, the main contribution to the strong phase 
in PQCD stems from the annihilation mechanism
with the scalar-pseudoscalar operator $O_6$, 
and  the diagrams again start at $O(\alpha_s)$ level.
We have also found $O_6$ to be an important source of 
factorizable annihilation,  but in LCSR the $O(\alpha_s)$ 
contribution to this mechanism is expected 
to be subleading  in comparison with the  zeroth order 
in $\alpha_s$,
``soft'' contribution which has been  calculated 
in sect.~\ref{Sec-O6}. 
Starting at $O(\alpha_s)$ level and neglecting soft
contributions, PQCD nevertheless 
predicts annihilation effects that are larger then 
in LCSR. This surprising fact means that 
it is difficult if not impossible to reconcile 
these two approaches with each other.

\section{Conclusion}
\label{Sec-Concl}

In this paper, the weak annihilation contributions to 
$B\to\pi\pi$ decay amplitudes have been calculated applying 
the method of LCSR. This work complements previous
studies \cite{AKBpipi,KMU,KMMpeng,KMMSU3} of nonfactorizable 
effects in $B\to\pi\pi$ with the QCD sum rule approach. 
In LCSR, due to sufficient virtuality 
of  the underlying correlation function, the OPE diagrams with 
annihilation topology  
are free from end-point divergences. Both contributions
of hard and soft gluons are taken into account.    
A finite result for the hadronic matrix element of 
the current-current $O_{1}^u$ operator 
with annihilation topology  is obtained 
including an imaginary part which contributes to the strong phase.
In addition, an important factorizable  
contribution from the quark-penguin operator $O_6$ 
has been  found. For the annihilation with hard gluons considered 
in this paper, we have modified the method suggested in
\cite{AKBpipi}, to avoid the problem of 
calculating two-loop multiscale
diagrams. Instead of performing the QCD calculation based on the 
vacuum-to-pion correlation function, we start from
the pion-pion correlator, thereby reducing the calculation to one-loop
diagrams.

We emphasize that QCD sum rules   
have a limited accuracy, at the same time one is able to 
estimate the uncertainties of the method. Moreover, many 
uncertainties  cancel in the ratios of nonfactorizable and 
factorizable hadronic matrix elements obtained from LCSR.
Obtaining  $B\to \pi\pi$ 
hadronic matrix elements  one 
uses an additional assumption of the 
local quark-hadron duality, allowing the transition
(analytical continuation) 
from a large spacelike scale to the large 
timelike scale $m_B^2$. This
approach has much in common with evaluating
the timelike asymptotics of the pion e.m. form factor 
from the QCD calculation in the spacelike region.  

Our main phenomenological result 
is a smallness of the annihilation contributions
in $B\to\pi\pi$. This is generally consistent with QCDF, 
if the divergent annihilation diagrams there are 
modelled by moderate logarithmic factors.  
Hence, we find  no compelling reason to consider
the annihilation amplitude
as a  free parameter in QCDF. The relatively large values
of this parameter generated by the fits to the data
are probably not originating from the annihilation mechanism.

Small annihilation effects predicted in this paper are 
in the same ballpark as other 
nonfactorizable contributions obtained from 
LCSR, including charming and gluonic penguin
topologies and nonfactorizable corrections
to the emission topology. Altogether, the smallness
of the corrections to the leading-order factorizable
amplitude reveals 
a good convergence of the OPE series for the correlation
function and justifies the use of the adopted approximation
in LCSR, that is, including only $O(\alpha_s)$ and twists $\leq 4$, 
as well as omitting the small $O(s_0^\pi/m_B^2)$ corrections 
in each term of OPE.

Furthermore, we have performed the phenomenological analysis 
of the three $B\to \pi\pi$ channels  using the results 
of LCSR for all nonfactorizable effects, except the hard-gluon
nonfactorizable corrections to the emission topology.
For the latter the default prediction of QCDF \cite{BBNS1,BBNS2} is used.
Our results disagree with the current data
for  $BR(B^0\to\pi^+\pi^-)$ and $BR(B^0\to\pi^0\pi^0)$ 
and probably also for the direct CP-asymmetry in 
$B^0\to\pi^+\pi^-$, that is, for the channels 
where the $\Delta I=1/2$ 
weak transition (or, equivalently, the $I=0$ two-pion final state) 
contributes. If the experimental data do not change, 
we have to admit that LCSR  misses an important part of 
the isoscalar amplitude. Then a new ``inverse $\Delta I=1/2$'' 
rule has to be established for these decays, meaning that the
amplitude $A_0$ introduced in sect.~2 has 
to decrease after including the missing piece.

A natural question arises: are there additional mechanisms
in $B\to\pi\pi$ which may fill this gap.
The penguin-annihilation topology comes first to one's mind, 
a possibility which was not yet explored 
by both QCDF and LCSR. This effect contains multiloop diagrams
and cannot be easily evaluated.
Still, the experience with LCSR for charming penguins
\cite{KMMpeng} tells us that 
the OPE diagrams of the type shown in Fig.~\ref{fig-OPE_PA} 
could not be large.
In any case, the penguin-annihilation mechanism 
deserves a closer look in future. 
One can for example speculate 
about nonperturbative gluonic effects
in this mechanism, which have an anomalously 
large scale ($\gg \Lambda_{QCD}$) and  lie beyond OPE. 
Such effects will most probably influence  
other  neutral final states, e.g., enhance  $B\to \rho^0\rho^0$.
That however seems not to be the case because 
the current experimental upper bound for $BR(B\to \rho^0\rho^0)$ 
is less than the measured value of $BR(B\to \pi^0\pi^0)$.

Another resource of enhancement is the hard-spectator part in 
the nonfactorizable emission. The corresponding
diagrams in QCDF diverge at twist 3 and are replaced by another
generic logarithm. In our numerical estimates  we have taken 
the default value of this parameter
from \cite{BBNS2}, but the fits to the current data \cite{BBNSfits} 
with a free parameter  
for this contribution produce large effects.
Indications that the hard-spectator 
mechanism is important, were found recently in the 
SCET framework \cite{BenekeYang}. We plan to study 
the nonfactorizable 
hard-gluon emission, including the hard-spectator 
mechanism in LCSR (work in progress).

Finally, an important avenue of future studies 
is the sum rule analysis of 
the $B\to \pi K$ and  $B\to K \bar{K}$ channels 
including calculable $SU(3)$ violation effects
which are expected \cite{KMMSU3} to be important. 
The variety of kaon channels with a lot   
of accumulated data will allow to isolate 
various topologies, and to put
the LCSR approach under a tighter scrutiny.

\section*{Acknowledgements} We are grateful to
Thorsten Feldmann for useful discussions and comments. 
This work is supported by DFG and the German Ministry of Education and Research (BMBF). 
The work of B.M. is partially 
supported by the Ministry of Science, Education and
Sport of the Republic of Croatia under contract No. 0098002.
\begin{appendix}
\section{Effective coefficients in QCDF}
\label{Sec-ri_vs_ai}

For convenience, we present here the relations
between the nonfactorizable matrix elements 
parameterized in section \ref{Sec-Pheno-nonfact} 
by the ratios $r_T^{(\pi\pi)}$  
and the effective coefficients $a_i$
introduced in \cite{BBNS1,BBNS2} for $B\to \pi\pi$.
The decay amplitudes written in terms of these 
coefficients are:
\ba
A(\bar{B}^0 \to \pi^+ \pi^-)&=&\Big[
\lambda_u a_1+ \sum\limits_{p=u,c}
\lambda_p (a_4^p+r_\chi^\pi a_6^p)\Big]
{\cal A}_{\pi\pi} \nonumber\\
&&+\Big[
\lambda_u b_1+ (\lambda_u +\lambda_c)
(b_3+ 2b_4)\Big]
{\cal B}_{\pi\pi}\,,
\nonumber\\ 
\sqrt{2}A(B^-\to \pi^-\pi^0)&=&\lambda_u(a_1+ a_2){\cal A}_{\pi\pi}\,,
\label{ABpipi1_ai}
\ea
where $r_\chi^\pi=2\mu_\pi/m_b$
and electroweak penguins are neglected.
Comparing with Eqs.(\ref{eq-decomptot}),
(\ref{eq-decomp1}),(\ref{eq-iso2Bpipi}),
(\ref{eq-iso0Bpipi}), (\ref{eq-rPc})
and (\ref{eq-3456Bpipi})
we obtain:
\ba
a_1=\,c_1+\frac{c_2}{3}+2\,c_2\,r_E^{(\pi\pi)},~~ 
a_2=\,c_2+\frac{c_1}{3}+2\,c_1\,r_E^{(\pi\pi)}\,,
\nonumber
\\
a_4^p+r_\chi a_6^p=
\,c_4+\frac{c_3}{3}+2\,c_3\,r_E^{(\pi\pi)}+
\,2\,c_1r_{P_p}^{(\pi\pi)}+
2\,c_3\left(r_{P_u}^{(\pi\pi)}+r_{P_b}^{(\pi\pi)}\right)
\nonumber \\
+2 (c_4+c_6)\left(3\bar{r}_{P_q}^{(\pi\pi)}+
\bar{r}_{P_c}^{(\pi\pi)}+\bar{r}_{P_b}^{(\pi\pi)}\right)
+r_\chi^\pi\left(c_6+\frac{c_5}{3}\right)
+ 2\,c_5\,r_E^{(\pi\pi,6)}+
c_{8g}^{eff}r_{8g}^{(\pi\pi)} \,,
\\
b_1= 2\,c_1\,r_A^{(\pi\pi)} 
\frac{{\cal A_{\pi\pi}}}{{\cal B_{\pi\pi}}}\,,
\quad
b_3= \Biggl[
2\,c_3\,r_A^{(\pi\pi)}+
\left(c_6+
\frac{c_5}{3}\right)R_A^{(\pi\pi,6)}+
2\,c_5\,r_A^{(\pi\pi,6)}\Biggr]
\frac{{\cal A_{\pi\pi}}}{{\cal B_{\pi\pi}}}\,,
\nonumber\\
b_4= 2\left[\,c_4\,r_A^{(\pi\pi)}+
\,c_6\,r_A^{(\pi\pi,5)}\right] 
\frac{{\cal A_{\pi\pi}}}{{\cal B_{\pi\pi}}}\,.
\ea
The correspondence between the annihilation 
diagrams introduced in \cite{BBNS2} 
and parameters $r_A^{(\pi\pi)}$ is
schematically given by 
\ba
r_A^{(\pi\pi)}{\cal A_{\pi\pi}}
\hat{=}\frac{C_F}{18}A_1^i{\cal B_{\pi\pi}}\,,~~
r_A^{(\pi\pi,5)}{\cal A_{\pi\pi}}
\hat{=}\frac{C_F}{18}A_2^i{\cal B_{\pi\pi}}\,,
\nonumber \\
R_A^{(\pi\pi,6)}{\cal A_{\pi\pi}}
\hat{=}\frac{C_F}{3}A_3^f{\cal B_{\pi\pi}}\,,\quad
r_A^{(\pi\pi,6)}{\cal A_{\pi\pi}}
\hat{=}\frac{C_F}{18}A_3^i
{\cal B_{\pi\pi}}\,,
\ea
where the sign $ \hat{=}$ indicates that 
in LCSR and QCDF different approximations
are used to calculate l.h.s and r.h.s., respectively.
In particular, \(R_A^{(\pi\pi,5)}\) is of the zeroth
order in \(\alpha_s\), whereas \(A_3^f\) is of 
\(O(\alpha_s)\).

\section{Kernels for hard annihilation}
\label{Sec-MessyFormula}
The kernels for the hard annihilation contribution to the correlation function
in Eq.~(\ref{eq-Pipipi-diagram}) are given by
\begin{equation*}
\label{eq-Hpipi}
\begin{split}
H_{\pi\pi}^d\left(u,v;s'=(p-q)^2,P^2\right)=
\Bigl[P^2\left(2 m_b^2-s'(2+u)+3 P^2 v\right)\Bigr]&B_0\left(P^2 u v,0,0\right)\\
-\Bigl[\frac{1}{u v}\left((s' u+P^2 v)m_b^2+2 P^4 v^2-s'^2 u-P^2 s' v\right)\Bigr]&B_0\left(s',0,m_b^2\right)\\
+\Bigl[\frac{1}{u v}\bigl((s' u+P^2 v (1-2 u))m_b^2+P^4 v^2(1-3 u)\\-s'^2 u(1+u)+vP^2 s'(u^2+4u-1)\bigr)\Bigr]
  & B_0\left((1-u)(s'-P^2 v),0,m_b^2\right)\\
+\Bigl[2 P^2\left(m_b^2-s'+P^2 v\right)^2\Bigr] C_0\bigl(s',P^2 u v,
&(1-u)(s'-P^2 v),m_b^2,0,0\bigr)\,,
\end{split}
\end{equation*}
\begin{equation*}
\begin{split}
H_{\pi\pi}^b\left(u,v;s'=(p-q)^2,P^2\right)=
-\Bigl[P^2\left(2 m_b^2+s'(2-3 u)+P^2 v\right)\Bigr]&B_0\left(P^2 u v,
m_b^2,m_b^2\right)\\
+\Bigl[\frac{1}{u v}\left((s' u+P^2 v)m_b^2+s'^2 u(1-2u)+s' P^2 v\right)\Bigr]
&B_0\left(s',0,m_b^2\right)\\
-\Bigl[\frac{1}{u v}\bigl((s' u+P^2 v (1-2 u))m_b^2-P^4 v^2(1+u)+s'^2 u(1-3
u)\\
+P^2 v s' (1+3 u^2)\bigr)\Bigr]&B_0\left((1-u)(s'-P^2 v),0,m_b^2\right)\\
+\frac{2}{uv}\Bigl[
(u m_b^2 + v P^2 - u^2 s')(v P^2 (m_b^2 - u s') + u s'^2)
\Bigr]C_0\bigl(s',P^2 u v,&(1-u)(s'-P^2 v),0,m_b^2,m_b^2\bigr)\,.
\end{split}
\end{equation*}
\(B_0\) and \(C_0\) are the standard two-point and three-point 
functions, respectively:
\begin{equation*}
B_0(p^2,m_0,m_1)=\frac{(2\pi\mu)^{4-D}}{i\,\pi^2}\int d^D q 
\left\{(q^2-m_0^2+i\epsilon)([q+p_1]^2-m_1^2+i\epsilon)\right\}^{-1}\,,
\end{equation*}
\begin{multline*}
C_0\left(p_1^2,(p_1-p_2)^2,p_2^2,m_0,m_1,m_2\right)\\
=\frac{(2\pi\mu)^{4-D}}{i\,\pi^2}\int d^D q
\left\{(q^2-m_0^2+i\epsilon)([q+p_1]^2-m_1^2+i\epsilon)
([q+p_2]^2-m_2^2+i\epsilon)\right\}^{-1}\,.
\end{multline*}

\end{appendix}

\end{document}